\documentclass{emulateapj} 
\usepackage{graphicx}

\newcommand{\be}{\begin{equation}}
\newcommand{\ee}{\end{equation}}
\newcommand{\ba}{\begin{eqnarray}}
\newcommand{\ea}{\end{eqnarray}}

\newcommand{\mic}{\mu{\rm m}}

\newcommand{\kms}{\rm km\,s^{-1}}

\newcommand{\Mpc}{\rm Mpc}

\def\ls{\mathrel{\hbox{\rlap{\hbox{\lower4pt\hbox{$\sim$}}}\hbox{$<$}}}}
\def\gs{\mathrel{\hbox{\rlap{\hbox{\lower4pt\hbox{$\sim$}}}\hbox{$>$}}}}
\shorttitle{The slow quenching of star formation in cluster galaxies}
\shortauthors{Haines et al.}
\begin{document}
\title{LoCuSS: The steady decline and slow quenching of star formation in cluster galaxies over the last four billion years}

\author{
  C.\ P.\ Haines,$\!$\altaffilmark{1,2,3}
  M.\ J.\ Pereira,$\!$\altaffilmark{1}
  G.\ P.\ Smith,$\!$\altaffilmark{2}
  E.\ Egami,$\!$\altaffilmark{1}
  A.\ J.\ R.\ Sanderson,$\!$\altaffilmark{2}
  A.\ Babul,$\!$\altaffilmark{4}
  A.\ Finoguenov,$\!$\altaffilmark{5,6}\\
  P.\ Merluzzi,$\!$\altaffilmark{7}
  G.\ Busarello,$\!$\altaffilmark{7}
  T.\,D.\ Rawle,$\!$\altaffilmark{1,8}
  N.\ Okabe,$\!$\altaffilmark{9}
  \\
}

\altaffiltext{1}{Steward Observatory, University of Arizona, 933 North
  Cherry Avenue, Tucson, AZ 85721, USA; cphaines@as.arizona.edu} 
\altaffiltext{2}{School of Physics and Astronomy, University of
  Birmingham, Edgbaston, Birmingham, B15 2TT, UK}
\altaffiltext{3}{Departamento de Astronom\'{i}a, Universidad de Chile, Casilla 36-D, Correo Central, Santiago, Chile}
\altaffiltext{4}{Department of Physics and Astronomy, University of Victoria, 3800 Finnerty Road, Victoria, BC, V8P 1A1, Canada}
\altaffiltext{5}{Department of Physics, University of Helsinki, Gustaf H\"{a}llstr\"{o}min katu 2a, FI-0014 Helsinki, Finland}
\altaffiltext{6}{Center for Space Science Technology, University of Maryland Baltimore County, 1000 Hilltop Circle, Baltimore, MD 21250, USA}
\altaffiltext{7}{INAF-Osservatorio Astronomico di Capodimonte, via Moiariello 16, I-80131 Napoli, Italy}
\altaffiltext{8}{European Space Astronomy Centre, ESA, Villanueva de la Ca\~{n}ada, E-28691 Madrid, Spain}
\altaffiltext{9}{Academia Sinica Institute of Astronomy and Astrophysics (ASIAA), P.O. Box 23-141, Taipei 10617, Taiwan}


\begin{abstract}
We present an analysis of the levels and evolution of star formation activity in a representative sample of 30 massive galaxy clusters at $0.15{<}z{<}0.30$ from the Local Cluster Substructure Survey (LoCuSS), combining wide-field {\em Spitzer}/MIPS 24$\mu$m data with extensive spectroscopy of cluster members. 
The specific-SFRs of massive ($\mathcal{M}{\ga}10^{10}{\rm M}_{\odot}$) star-forming cluster galaxies within $r_{200}$ are found to be systematically ${\sim}28$\% lower than their counterparts in the field at fixed stellar mass and redshift, a difference significant at the $8.7{\sigma}$ level. This is the unambiguous signature of star formation in most (and possibly all) massive star-forming galaxies being {\em slowly} quenched upon accretion into massive clusters, their star formation rates (SFRs) declining exponentially on quenching time-scales in the range 0.7--2.0\,Gyr. 
We measure the mid-infrared Butcher--Oemler effect over the redshift range \mbox{0.0--0.4}, finding rapid evolution in the fraction ($f_{SF}$) of massive ($M_{K}{<}{-}23.1$) cluster galaxies within $r_{200}$ with SFRs${>}3$\,M$_{\odot}$\,yr$^{-1}$, of the form $f_{SF}{\propto}(1+z)^{7.6{\pm}1.1}$. We dissect the origins of the Butcher--Oemler effect, revealing it to be due to the combination of a ${\sim}3{\times}$ decline in the mean specific-SFRs of star-forming cluster galaxies since $z{\sim}0.3$ with a ${\sim}1.5{\times}$ decrease in number density. Two-thirds of this reduction in the specific-SFRs of star-forming cluster galaxies is due to the steady cosmic decline in the specific-SFRs among those field galaxies accreted into the clusters. The remaining one-third reflects an {\em accelerated} decline in the star formation activity of galaxies within clusters. 
The slow quenching of star-formation in cluster galaxies is consistent with a gradual shut down of star formation in infalling spiral galaxies as they interact with the intra-cluster medium via ram-pressure stripping or starvation mechanisms.
The observed sharp decline in star formation activity among cluster galaxies since $z{\sim}0.4$ likely reflects the increased susceptability of low-redshift spiral galaxies to gas removal mechanisms as their gas surface densities decrease with time. 
We find no evidence for the build-up of cluster S0 bulges via major nuclear star-burst episodes. 
\end{abstract}

\keywords{ galaxies: spiral --- galaxies: clusters: general ---
  galaxies: evolution}

\section{Introduction}
\label{intro}

\setcounter{footnote}{9}

In the current concordance $\Lambda$CDM cosmological models, structure formation occurs hierarchically, with low-mass dark matter (DM) halos forming first at high redshifts, and more massive halos forming later via the merging, or accretion, of these low-mass halos. In this framework, the most massive collapsed objects at any epoch are the dynamically most immature. In the local Universe these most massive, immature DM halos host galaxy clusters. 
This hierarchical framework of structure formation affects galaxy evolution in two ways. 

Firstly, the massive {\em central} galaxies of DM halos are themselves hierarchically assembled via the merger and accretion of smaller galaxies, a process which primarily occurs at high redshifts. Numerical simulations show that massive early-type galaxies can be assembled rapidly in this ``revised-monolithic'' scheme \citep{merlin}, consuming their gas in a short burst, before being quenched either via extreme QSO feedback triggered by major mergers \citep[e.g.][]{springel,hopkins06,hopkins08}, or quasi-continual ``radio feedback'' from low-luminosity AGN \citep[e.g.][]{croton,bower}. The extent of this process depends on the initial conditions in which galaxies form. Massive galaxies are formed earlier in the highest overdensities of the primordial density field \citep{maulbetsch}, and have a more active merger history at $z{\ga}1$ \citep{gottlober}, than those that form in the smoother low-density regions. This leaves its imprints in the environmental trends of galaxy poplations observed at early epochs \citep[e.g.][]{cooper,quadri07,muzzin}. 

Secondly, as the DM halos which host many of these massive {\em central} galaxies are accreted onto more massive systems, these galaxies become {\em satellites} within galaxy groups and clusters. This process of becoming satellites is also known to affect the ability of galaxies to continue forming stars, as empirically the fraction of satellite galaxies which have become passive is always greater than for central galaxies of the same stellar mass \citep[e.g.][]{weinmann,wetzel}. 

The most plausible and likely causes of the systematic quenching of star formation in galaxies upon accretion into massive halos are their interactions with the dense intra-cluster medium (ICM) which pervades the virialized regions of groups and clusters. This occurs either via the partial or complete ram-pressure stripping of their cold gas contents due to their high-velocity (${\ga}1$\,000\,km\,s$^{-1}$) passage through the dense ICM \citep{gunn}, or the milder starvation mechanism in which they are simply prevented from accreting new gas from the surrounding inter-galactic medium, and so slowly consume their remaining cold gas through star formation over a period of Gyr \citep{larson,mccarthy}. Observations of spirals in local clusters show evidence for ongoing ram-pressure stripping, with 10--30\,kpc long one-sided extra-galactic tails of H{\sc i} gas all pointing away from the cluster center \citep{chung}, similar to those produced in hydrodynamical simulations \citep{tonnesen}. Many cluster spirals also show truncated H{\sc i} radial profiles and H$\alpha$ disks characteristic of the selective outside-in removal of gas symptomatic of ram-pressure stripping \citep{koopmann,vangorkom}. 

The key to establishing which physical processes are responsible for quenching star formation in galaxies when they are accreted into massive halos, as well as quantifying their efficiency, is to derive empirical constraints for the average time-scales required by the quenching mechanism to shut down star formation in these satellite galaxies, the dependence on halo mass for its efficiency, and its effective range in terms of cluster-centric radius \citep[e.g.][]{treu,boselli}. However, to date it remains unresolved whether this environmental quenching process is a rapid \citep[${\tau}{\la}1$00\,Myr;][]{balogh04,haines07,mcgee11} or slow \citep[${\ga}1$\,Gyr;][]{moran07,vonderlinden,lu} process, or whether it includes a prior starburst phase \citep{moran05,li}. There is also disagreement as to whether environmental quenching becomes more efficient with increasing host DM halo mass \citep{weinmann,wetzel11}, or is alternatively independent of halo mass \citep{balogh10,peng12}. 

The combined impacts of these two aspects of structure formation are responsible for producing the strong environmental trends observed in local galaxy populations, as classically quantified by the star formation (SF)--density \citep{dressler85} and morphology--density relations \citep{dressler80}. Field galaxies that have evolved in relative isolation are primarily star-forming spirals, while moving to the densest regions in the cores of groups and clusters, the dominant population shifts to passive ellipticals and S0s, with star-forming spirals now essentially absent. 

Cluster galaxies have not always been as inactive as they are at the present epoch. \citet[hereafter BO84]{bo84} showed that the fraction of blue (star-forming) galaxies among cluster members increases from almost zero in the local Universe to ${\sim}20$\% by $z{\sim}0.4$. This implies a rapid evolution in the cluster population over the last five billion years, in which empirically the star-forming spiral galaxies found by BO84 in clusters at $z{\sim}0.4$ are mostly replaced by passive S0s in local clusters \citep{dressler97,treu}. More recently, infrared studies of cluster galaxies with {\em Spitzer} \citep{werner} and {\em Herschel} \citep{pilbratt} have mapped in detail this rapid increase of star formation activity among cluster galaxies to $z{\sim}1$ and beyond \citep[e.g.][]{saintonge,finn10,koyama,tran10,popesso12}. While LIRGs (${\rm SFRs}{\ga}1$0\,M$_{\odot}$\,yr$^{-1}$) have been shown to be essentially absent in all {\em Spitzer} analyses of nearby ($z{<}0.1$) rich clusters \citep{bai06,bai09,haines11a}, dusty star-forming galaxies with ${\rm SFRs}{\ga}3$0\,M$_{\odot}$\,yr$^{-1}$ were found to be common in clusters at $z{\sim}0.$4--0.5 \citep{geach,geach09,marcillac,oemler}.

The systematic decline in star formation in galaxies since $z{\sim}1$ is however not just limited to those in clusters, but has been observed in all environments. Both the {\em cosmic} star formation rate (SFR) and the median SFRs of star-forming field galaxies (at fixed stellar mass) has declined by ${\sim}10{\times}$ since $z{\sim}1$ \citep{lefloch,zheng,oliver}. As the SFRs of essentially all star-forming galaxies have been declining in unison (for a given stellar mass) since $z{\sim}$1--2, with little dependence on environment, this must be driven by processes inherent to galaxies as a population. 
\citet{dutton} and \citet{dave} have argued that this cosmic decline in star formation is driven primarily by the parallel fall in the cosmological rate of baryonic accretion onto dark matter halos, which scales as $(1+z)^{2.25}$ at fixed halo mass \citep{birnboim}. 

Clusters and their member galaxies do not exist and evolve in isolation. Clusters are located at the nodes of the filamentary network which makes up the large-scale structure of the Universe, and have been continually accreting mass in the form of galaxies and galaxy groups from their surroundings. \citet{berrier} show that the accretion rate of galaxies onto clusters from their surroundings has been fairly uniform over the last 10\,Gyr, the median time since accretion for a galaxy in a cluster at the present day being $\sim$4--5\,Gyr. \citet{mcgee09} indicate that ${\sim}4$0--50\% of those galaxies accreted into massive clusters (${\sim}10^{15}\,{\rm M}_{\odot}$) since $z{\sim}0.5$ arrived as members of galaxy groups rather than as isolated field galaxies. This implies that a significant number of star-forming spirals have been infalling onto clusters at late epochs ($z{\la}0.4$). These must then be rapidly transformed by environmental processes, either in-situ or pre-processed within galaxy groups, in order to leave the present day clusters as devoid of star forming galaxies as observed.
This late and continual accretion of galaxies into clusters leaves its imprint in the form of strong radial population gradients in the accretion epochs of galaxies. These can be modelled with cosmological simulations and have been successfully tuned to qualitatively reproduce the observed SF--density relation \citep[e.g.][]{balogh,ellingson} and constrain the time-scale required to transform galaxies \citep[e.g.][hereafter McG09, McG11]{smith12,mcgee09,mcgee11}. 

These issues have motivated us to embark on a multi-wavelength survey, centered around wide-field {\em Spitzer} and {\em Herschel} infrared (24--500$\mu$m) imaging of 30 massive clusters at $0.15{\le}z{<}0.30$ selected from the Local Cluster Substructure Survey (LoCuSS\footnote{http://www.sr.bham.ac.uk/locuss/}$^{,}$\footnote{http://herschel.as.arizona.edu/locuss/locuss.html}). Such a large statistical sample of clusters is mandatory to handle the large cluster-to-cluster scatter seen in their galaxy populations (e.g. BO84), resulting from the wide variations in the dynamical states and current mass accretion rates of massive clusters \citep{smith08}, a reflection of their dynamical immaturity. 
The different physical processes believed to contribute to the transformation of cluster galaxies act over a wide range of cluster-centric radii \citep{treu,smith10b}, necessitating panoramic datasets extending to the virial radius and beyond to provide any discriminatory power.  Moreover, mapping the large-scale structure in which each cluster is embedded is vital to understand how galaxy evolution is dependent on the hierarchical assembly of massive clusters.  The overall science goals and early results from the {\em Herschel} data are presented in \citet{smith10b}, \citet{haines10} and \citet{pereira10}.

In \citet[hereafter Paper I]{haines09a} we combined panoramic {\em Spitzer}/MIPS 24$\mu$m imaging of 22 clusters from LoCuSS with comparable data for eight more systems to measure the evolution with redshift of $f_{SF}$, the fraction of massive infrared-luminous cluster galaxies ($M_{K}{<}M_{K}^{*}{+}1.5$, $L_{IR}{>}5{\times}10^{10}L_{\odot}$) within $r_{200}$ --- the mid-infrared (MIR) Butcher-Oemler effect. We found rapid evolution of the form $f_{SF}{\propto}(1+z)^{5.7}$, paralleling that seen in the original BO84 study. We then showed that this apparent rapid evolution was due primarily to the {\em cosmic} decline in star formation in field galaxies, which were accreted onto clusters at a constant rate and subsequently quenched by cluster related processes. 

The key limitation of Paper I was that it was based on partial redshift information, coming primarily from the literature (and hence varying dramatically from cluster to cluster), such that we had to rely on statistically accounting for the contribution from field galaxy interlopers when estimating the $f_{SF}$. We have since completed ACReS (the Arizona Cluster Redshift Survey\footnote{http://herschel.as.arizona.edu/acres/acres.html}) a long-term spectroscopic program to observe our sample of 30 clusters with MMT/Hectospec, providing redshifts for ${\sim}10$\,000 cluster members. 

In this paper we combine the panoramic {\em Spitzer} imaging with this extensive redshift data to obtain a complete census of obscured star-formation among cluster galaxies for a statistical sample of 30 clusters at $0.15{\le}z{<}0.30$, extending into the infall regions beyond the virial radius. 
These data allow us to firstly robustly measure the MIR Butcher--Oemler effect (\S\ref{sec:bo}), but then also dissect its origins, quantifying the relative contributions from the steady decline in the specific-SFRs of star-forming galaxies in clusters as well as reductions in their numbers over the last 4\,Gyr. 
In {\S}\ref{sec:ssfr} we directly compare the specific-SFRs of star-forming galaxies in clusters and their field counterparts at fixed stellar mass and redshift, finding them to be systematically lower in clusters. This provides unambiguous evidence that star formation in galaxies is slowly quenched as they encounter the cluster environment for the first time, and in  {\S}\ref{sec:slowquench} we estimate a quenching time-scale in the range 0.7--2.0\,Gyr, by comparison with predictions of the continual accretion of galaxies onto rich clusters from cosmological simulations. 
Throughout we assume \mbox{$\Omega_M{=}0.3$}, \mbox{$\Omega_\Lambda{=}0.7$} and \mbox{${\rm H}_0{=}70\,\kms\Mpc^{-1}$}.

\section{Sample and Observations}
\label{sec:data}

\subsection{Sample}

LoCuSS is a multi-wavelength survey of a morphologically unbiased sample of ${\sim}8$0 X-ray luminous galaxy clusters at $0.15{\le}z{\le}0.3$ \citep{smith10a} drawn from the ROSAT All Sky Survey cluster catalogs \citep{ebeling98,ebeling00,bohringer}. The first batch of 30 clusters in our survey (Table 1) benefits from a particularly rich dataset, including: Subaru/Suprime-Cam optical imaging \citep{okabe10}, {\em Spitzer}/MIPS $24\mic$ maps, {\em Herschel}/PACS+SPIRE 100--500$\mic$ maps, {\em Chandra} and/or {\em XMM} X-ray data, and near-infrared (NIR; $J,K$) imaging.  All of these data embrace at least $25^{\prime}{\times}25^{\prime}$ fields of view centered on each cluster,  and thus probe the clusters out to ${\sim}1$.5--2 virial radii \citep{smith10b}. These 30 clusters were selected from the parent sample simply on the basis of being observable by Subaru on the nights allocated to us \citep{okabe10}, and should therefore not suffer any gross biases towards (for example) cool core clusters, merging clusters etc. Indeed, \citet{okabe10} show that the sample is statistically indistinguishable from a volume-limited sample.

\subsection{Mid-infrared Observations}

Each cluster was observed across a $25^\prime{\times}25^\prime$ field of view at $24\mic$ with the MIPS instrument \citep{rieke04} on board the {\em Spitzer Space Telescope}\footnote{This work is based in part on observations made with the Spitzer Space Telescope, which is operated by the Jet Propulsion Laboratory, California Institute of Technology under a contract with NASA (contract 1407).}, consisting of a $5{\times}5$ grid of MIPS pointings in fixed cluster or raster mode (PID: 40872; PI: G.P. Smith). At each grid point we performed two cycles of the small-field photometry observations with a frame time of 3s, producing a total per pixel exposure time of 90s. The $24\mic$ mosaics were analyzed with SExtractor \citep{bertin} producing catalogs which are on average 90\% complete to 400$\mu$Jy. Details of the reduction, source extraction and photometry can be found in \citet{haines09a}.

\subsection{Near-infrared Observations}

$J$- and $K$-band near-infrared images of 26 of the 30 clusters were obtained with WFCAM on the 3.8-m United Kingdom Infrared Telescope (UKIRT)\footnote{UKIRT is operated by the Joint Astronomy Centre on behalf of the Science and Technology Facilities Council of the United Kingdom.} in service mode over multiple semesters starting in March 2008.
 The WFCAM data cover $52'{\times}52'$ fields to depths of $K{\sim}19$, $J{\sim}21$ (Vega) with exposure times of 640s per pointing, pixel size of $0.2''$ and ${\rm FWHMs}{\sim}0$.7--1.$2''$. 
The remaining four clusters were observed with NEWFIRM on the 4.0-m Mayall telescope at Kitt Peak\footnote{Kitt Peak National Observatory, National Optical Astronomy Observatory, which is operated by the Association of Universities for Research in Astronomy (AURA) under cooperative agreement with the National Science Foundation.} on 17 May 2008 and 28 December 2008. 
The NEWFIRM data consist of dithered and stacked $J$- and $K$-band images covering $27'{\times}27'$ fields of view with a $0.4''$ pixel-scale and ${\rm FWHM}{\sim}1$.0--1$.5''$. The total exposure times in each filter were 1800s, and also reach depths of $K{\sim}19$, $J{\sim}21$.

Total $J$- and $K$-band Kron magnitudes were determined for each source, while $J{-}K$ colors were derived within fixed circular apertures of diameter 2\,arcsec. 

\subsection{MMT/Hectospec spectroscopy}
\label{spectroscopy}

We have recently completed ACReS (the Arizona Cluster Redshift Survey; Pereira et al. 2013 in preparation) a long-term spectroscopic program to observe our sample of 30 galaxy clusters with MMT/Hectospec. Hectospec is a 300-fiber multi-object spectrograph with a circular field of view of $1^{\circ}$ diameter \citep{fabricant} on the 6.5m MMT telescope at Mount Hopkins, Arizona. This telescope and instrument combination is ideal for studying galaxies within and in the vicinity of massive clusters at these redshifts, extending to cluster-centric radii of 5--7.5\,Mpc, and matching well our panoramic near-infrared coverage from WFCAM. We used the 270 line grating, which provides a wide wavelength range (3650--9200{\AA}) at 6.2{\AA} resolution. 
Redshifts were determined by comparison of the reduced spectra with stellar, galaxy and quasar template spectra, choosing the template and redshift which minimizes the $\chi^{2}$ between model and data. 

Following Paper I, probable cluster members were identified from the $JK$ photometry, based on the empirical observation that galaxies of a particular redshift lie along a single narrow $J{-}K/K$ color-magnitude relation of width $\sigma_{J-K}{\sim}0$.04--0.06\,mag (see their Fig~1).  This relation evolves monotonically redward with redshift to $z{\simeq}0.5$ \citep[see][]{haines09b}.  The NIR colors of galaxies are relatively insensitive to star-formation history and dust extinction, with the $J{-}K$ color varying by only ${\sim}0.1{\rm mag}$ across the entire Hubble sequence, and hence only a single sequence is seen, unlike in the optical where separate red and blue sequences are visible. 

For each cluster, the location of the single C-M relation in $J{-}K/K$-space was identified using a combination of the existing spectroscopic members from the literature (where available) and/or photometric galaxies in the core of the cluster, where the contrast of the cluster sequence over the background field population should be greatest. Having identified the main sequence, and quantified the slope of the C-M relation, target galaxies were simply selected as lying within a color slice of width 0.3--0.4\,mag enclosing the observed cluster sequence, whose color cuts were fixed parallel to the C-M relation. 
While the lower limit was placed immediately below the sequence, the upper limit was generally fixed ${\sim}0$.25 mag redder in $J{-}K$ than the middle of the sequence to include the dust-obscured cluster population, many of which lie slightly above the sequence (see Fig.~1 of Paper I). 

The optimal strategy for positioning the color cuts with respect to the C-M relation was determined by examining of the C-M diagrams of spectroscopic members of eight clusters which already had extensive redshifts (60--200 per cluster) from the literature (e.g. A1689, A1835; see Fig.~1 of Paper I). The final color cuts were found to retain 98\% of spectroscopic members from these eight clusters, while reducing the number of galaxies to be targetted for spectroscopy by a factor 2. Many of the 2\% of cluster galaxies missed by the color cuts have peculiar $J{-}K$ colors due to a superposition of the galaxy with either a star or background galaxy, and overall they show no bias towards any particular galaxy class or star-formation history.

The primary aims of the redshift survey were to produce unbiased, stellar-mass limited samples of cluster galaxies with ${\ga}200$ members per cluster, suitable for galaxy evolution studies, deriving robust dynamical mass estimates and mapping in detail the surrounding large-scale structure. At the same time we wished to obtain a virtually complete census of obscured star formation in the cluster population and maximize the redshift completeness for sources detected in our {\em Spitzer} 24$\mu$m and {\em Herschel} 100--500$\mu$m images.  Given the above stated science goals, our targeting strategy for placing fibers was in decreasing order of priority:
\begin{enumerate}
\item All sources with $f_{24}{>}1$\,mJy in the galaxy catalog, irrespective of color. This ensures complete redshift coverage of cluster LIRGs, including those powered by AGN.

\item All probable cluster members (based on their $J{-}K$ color) with $0.4{<}f_{24}{<}1.0$\,mJy. This aims to complete the census of obscured star-formation within cluster galaxies down to SFRs of 1--3\,M$_{\odot}{\rm yr}^{-1}$.

\item Probable cluster galaxies with $M_{K}{<}M_{K}^{*}{+}2.0$, prioritizing firstly those galaxies within the cluster core, where it is more difficult to place fibers, and those brighter than $L_{K}^{*}$. 
\end{enumerate}

Our spectroscopic survey should also show no morphological bias against unresolved extragalactic objects such as compact galaxies or QSOs at the cluster redshift. This is because we have targeted sources based solely upon their $J{-}K$ colors or 24$\mu$m emission, irrespective of whether they are resolved or not in our near-IR images. Stars show much bluer near-IR colors ($J{-}K{\la}1$) than galaxies or quasars at the redshift of interest \citep[see Figs 2, 3 of][]{haines09b}, and were identified and excluded as targets for spectroscopy {\em only} if they had $J{-}K{<}1$ and were unresolved in our $K$-band data. 

\begin{table*}
\centering
\begin{tabular}{lcr@{:}llr@{ }lcr@{.}lr@{.}lcc} \hline
Cluster   & Right Ascension & \multicolumn{2}{c}{Declination} & \multicolumn{1}{c}{${<}z{>}$} &\multicolumn{2}{c}{$N_{z}$ (lit)}&$C_{z}$(ACReS) &\multicolumn{2}{c}{$r_{500}$}&\multicolumn{2}{c}{$r_{200}$}& $M_{200}$ & $f_{SF}$ \\
Name          &  (J2000) & \multicolumn{2}{c}{(J2000)}    &       &  &        & ($f_{24}{>}0.4$\,mJy) & \multicolumn{2}{c}{(Mpc)} & \multicolumn{2}{c}{(Mpc)} & ($10^{14}{\rm M}_{\odot}$) & \\ \hline 
Abell 68      & 00:37:06.84 & +09:09&24.3 & 0.2510 & 194&(0)     & 127/130 & 0&955 & 1&394  & 3.101 & $0.069_{-0.027}^{+0.038}$\\
Abell 115    & 00:55:50.65 & +26:24&38.7 & 0.1919 & 213&(36)   &   77/96   & 1&304 & 1&673  & 5.346 & $0.026_{-0.013}^{+0.020}$\\
Abell 209    & 01:31:53.45 & -13:36&47.8 & 0.2092 & 393&(49)   & 104/111 & 1&230 & 1&842  & 7.140 & $0.063_{-0.020}^{+0.028}$\\
Abell 267    & 01:52:48.72 & +01:01&08.4 & 0.2275 & 88&(1)      & 148/153 & 0&994 & 2&839  & 26.16 & $0.114_{-0.041}^{+0.052}$\\
Abell 291    & 02:01:43.11 & -02:11&48.1 & 0.1955 & 126&(0)     &   99/101 & 0&868$^{a}$& 1&315& 2.598 & $0.042_{-0.023}^{+0.040}$\\
Abell 383    & 02:48:03.42 & -03:31&45.1 & 0.1887 & 175&(1)     & 171/192 & 1&049 & 1&516  & 3.977 & $0.111_{-0.035}^{+0.047}$\\
Abell 586    & 07:32:20.42 & +31:37&58.8 & 0.1707 & 247&(21)   & 154/159 & 1&150 & 1&634  & 4.978 & $0.042_{-0.015}^{+0.021}$\\
Abell 611    & 08:00:55.92 & +36:03&39.6 & 0.2864  & 297&(7)    & 197/211 & 1&372 & 2&029  & 9.583 & $0.084_{-0.022}^{+0.028}$\\
Abell 665    & 08:30:57.36 & +65:51&14.4 & 0.1827 & 359&(31)   & 129/129 & 1&381 & 2&193  & 12.04 & $0.083_{-0.019}^{+0.023}$\\
Abell 689    & 08:37:24.57 & +14:58&21.1 & 0.2786  & 186&(1)    & 157/163 & 1&126$^{b}$& 1&706& 5.688 & $0.084_{-0.024}^{+0.030}$\\
Abell 697    & 08:42:57.69 & +36:21&58.5 & 0.2818 & 377&(32)   & 165/175 & 1&505 & 1&935  & 8.309 & $0.098_{-0.020}^{+0.024}$\\
Abell 963    & 10:17:01.20 & +39:01&44.4 & 0.2043 & 425&(9)     & 420/470 & 1&275 & 1&808  & 6.750 & $0.122_{-0.929}^{+0.034}$\\
Abell 1689  & 13:11:29.45 & -01:20&28.3 & 0.1851 & 797&(356) & 149/150 & 1&501 & 2&129  & 11.01 & $0.055_{-0.016}^{+0.020}$\\
Abell 1758  & 13:32:44.47 & +50:32&30.5 & 0.2775 & 426&(5)     & 323/325 & 1&376 & 2&778  & 24.58 & $0.089_{-0.018}^{+0.022}$\\
Abell 1763  & 13:35:16.32 & +40:59&45.6 & 0.2323 & 303&(6)     & 505/510 & 1&220 & 1&845  & 7.182 & $0.046_{-0.014}^{+0.019}$\\
Abell 1835  & 14:01:02.40 & +02:52&55.2 & 0.2520 & 1077&(602)& 199/203 & 1&589 & 2&267  & 13.34 & $0.093_{-0.018}^{+0.020}$\\
Abell 1914  & 14:25:59.78 & +37:49&29.1 & 0.1671 & 394&(5)     & 190/192 & 1&560 & 1&889  & 7.690 & $0.038_{-0.014}^{+0.020}$\\
Abell 2218  & 16:35:52.80 & +66:12&50.4 & 0.1733 & 342&(49)   &   85/86   & 1&258 & 1&824  & 6.924 & $0.032_{-0.013}^{+0.018}$\\
Abell 2219  & 16:40:22.56 & +46:42&21.6 & 0.2257 & 415&(84)   & 208/221 & 1&494 & 2&770  & 24.30 & $0.061_{-0.015}^{+0.017}$\\
Abell 2345  & 21:27:13.73 & -12:09&46.1 & 0.1781 & 405&(39)   & 127/127 & 1&249$^{a}$& 1&893& 7.743 & $0.019_{-0.010}^{+0.018}$\\
Abell 2390  & 21:53:36.72 & +17:41&31.2 & 0.2291 & 517&(140) & 211/211 & 1&503 & 2&211  & 12.36 & $0.056_{-0.014}^{+0.019}$\\
Abell 2485  & 22:48:31.13 & -16:06&25.6 & 0.2476 & 196&(0)     & 134/134   & 0&830 & 1&262  & 2.300 & $0.000_{-0.000}^{+0.026}$\\
RXJ0142.0+2131& 01:42:02.64 & +21:31&19.2 & 0.2771&204&(15)&131/138 & 1&136 & 1&705  & 5.683 & $0.117_{-0.031}^{+0.037}$\\
RXJ1720.1+2638& 17:20:10.14 & +26:37&30.9 & 0.1599&363&(2) & 107/107 & 1&530 & 2&020  & 9.402 & $0.055_{-0.020}^{+0.029}$\\
RXJ2129.6+0005& 21:29:40.02 & +00:05&20.9 & 0.2337&259&(3) & 164/165 & 1&227 & 1&726  & 5.880 & $0.066_{-0.026}^{+0.037}$\\
ZwCl0104.4+0048&01:06:49.50&+01:03&22.1 & 0.2526 & 185&(1)& 136/141 & 0&760$^{a}$& 1&152 & 1.748 & $0.023_{-0.019}^{+0.051}$\\
ZwCl0823.2+0425&08:25:57.84&+04:14&47.5 & 0.2261 & 337&(4)& 181/192 & 1&050 & 1&582  & 4.526 & $0.028_{-0.018}^{+0.037}$\\
ZwCl0839.9+2937&08:42:56.06&+29:27&25.7 & 0.1931 & 173&(3)& 111/111 & 1&107 & 2&335  & 14.53 & $0.070_{-0.030}^{+0.045}$\\
ZwCl0857.9+2107&09:00:36.86&+20:53&40.0 & 0.2344 & 147&(0)& 109/111 & 1&024 & 1&519  & 4.009 & $0.098_{-0.047}^{+0.070}$\\
ZwCl1454.8+2233&14:57:15.23&+22:20&34.0 & 0.2565 & 157&(1)& 227/227 & 1&128 & 1&565  & 4.389 & $0.027_{-0.018}^{+0.034}$\\ \hline
\end{tabular}
\caption{The cluster sample. Column (1) Cluster name; cols. (2,3) Right Ascension, Declination of cluster center (J2000); col. (4) Mean redshift of cluster members; col. (5) Total number of spectroscopic cluster members (contribution taken from the literature); col. (6) Spectroscopic completeness of possible cluster members with $f_{24}{>}0$.4mJy; col. (7) radius $r_{500}$ in Mpc. $^{a}$From Martino et al. (2013), $^{b}$From Giles et al. (2012); col. (8) radius $r_{200}$ in Mpc; col. (9) Cluster mass $M_{200}$ in $10^{14}{\rm M}_{\odot}$; col. (10) $f_{SF} (r_{proj}{<}1.5\,r_{500})$.}
\label{clusterlist}
\end{table*}

For each cluster, we filled three to six configurations with targets, placing typically 200--250 of the 300 fibers on galaxies. 
Over the full survey of 30 systems, we reached overall completeness levels of 70\% for the $M_{K}$-selected sample, increasing to 96.4\% (5245/5441) for those galaxies detected with {\em Spitzer}. In the case of some of the highest redshift clusters, the sheer number of targets meant that we had to limit the $K$-band sample to $M_{K}{<}M_{K}^{*}{+}1.5$ rather than $M_{K}^{*}{+}2.0$.

To date, ACReS has required the equivalent of 13 full nights of observations since December 2008, producing ${\sim}3$0\,000 spectra, of which ${\sim}1$0\,000 have been identified as cluster members. For each cluster we typically obtained redshifts for 150--500 cluster members, the number depending primarily on the richness and/or compactness of the cluster (see Column~5 of Table 1). 

\subsection{Chandra/XMM X-ray imaging}

All but two (Abell 2345 and Abell 291) of the 30 clusters have available deep {\em Chandra} data. The observations of 21/28 clusters were made with the I mode of the Advanced Camera for Imaging Spectroscopy (ACIS-I) which has a field of view of $16.9^{\prime}{\times}16.9^{\prime}$, the remaining seven being observed with the smaller ACIS-S ($8.3^{\prime}{\times}8.3^{\prime}$). Total exposure times were in the range 9--120\,ksec. 

The deprojected dark matter densities, gas densities and temperature profiles for each cluster were derived by fitting the phenomenological cluster models of \citet{ascasibar} to a series of annular spectra extracted for each cluster \citep{sanderson10}. The best-fitting cluster models were then used to estimate $r_{500}$ and $r_{200}$, the radii inside which the enclosed densities are 500 and 200 times the critical density of the Universe at the cluster redshift \citep{sanderson09}. The X-ray emission for Abell 689 is dominated by a central BL Lac, and so we use the $r_{500}$ value from \citet{giles} who separated the extended cluster X-ray emission from the central point source. For the two clusters lacking {\em Chandra} data, we use the $r_{500}$ values derived by Martino et al. (2013) from {\em XMM} observations. 

 The {\em Chandra} data were also used to identify X-ray AGN as described in \citet{haines12}, using the wavelet-detection algorithm {\sc ciao wavdetect}. The survey limit of six broad band (0.3--7\,keV) X-ray counts results in on-axis sensitivity limits of $L_{X}{\le}1.0{\times}10^{42}\,{\rm erg\,s}^{-1}$ for X-ray AGN at the cluster redshift for all 28 systems \citep[see Table 1 from][]{haines12}.

Twenty-four of the 30 clusters have available deep {\em XMM} data, which we use to identify other galaxy groups and clusters in the region. Each 0.5--2\,keV {\em XMM} image is decomposed into unresolved and extended emission, following the wavelet technique of \citet{finoguenov}. For each extended source, we attempt to identify the redshift of its associated group/cluster by examining the Subaru optical images for likely BCGs near the center of the X-ray emission and/or groups of galaxies with similar redshifts from ACReS within the X-ray contours.

\subsection{Non-LoCuSS clusters}

To extend the redshift range of our Butcher-Oemler analysis outside the $0.15{<}z{<}0.30$ of the LoCuSS sample, we have analysed another 9 massive clusters with comparably wide mid-infrared coverage and extensive redshift data: the Coma cluster at $z{=}0.023$ \citep{bai06}; the 5 clusters forming the core of the Shapley supercluster (A\,3556, A\,3558, A\,3562, SC\,1329-313 and SC\,1327-312) at $z{=}0.048$ covered by the ACCESS\footnote{http://www.na.astro.it/ACCESS/} survey \citep{haines11a,haines11b,haines11c}; Abell 3266 at $z{=}0.060$ \citep{bai09}; Cl\,0024+17 at $z{=}0.394$ \citep{moran07,geach}; and Abell 851 at $z{=}0.405$ \citep{oemler}.

\section{Derived properties and final merged cluster and field galaxy catalogs}
\label{sec:catalogs}

\subsection{Identification of cluster members}

 For each cluster, we plot redshift against cluster-centric radius, identifying cluster members as lying with the general ``trumpet''-shaped caustic profile expected for galaxies infalling and subsequently orbiting within a massive virialized structure \citep{dunner}. 
\citet{rines06,rines} find that all massive clusters show clean infall patterns, with little ambiguity in the location of the caustics or limit of the pattern in redshift-space, and for most of our systems, there is a strong contrast in phase-space density from inside to outside these caustics, making their visual identification relatively simple.  

In Table~\ref{clusterlist} we list the 30 clusters in our sample, along with their central redshifts and $r_{500}$ values. We report the total number of spectroscopic members over the full region covered by ACReS, with the contribution from redshifts obtained from the literature (using NED) indicated in parentheses.

\subsection{A control sample of field galaxies at $0.15{<}{z}{<}0.30$}
\label{sec:field}

 To quantify the impact of the cluster environment on star formation in galaxies as they are accreted from the surrounding field, it is vital to have a detailed census of star formation among field galaxies at the same redshifts. 
It is also mandatory that the comparison field galaxies used are selected in an {\em identical} manner to those of the cluster. As the SFR of a galaxy depends not only on environment, but also its stellar mass and evolution with redshift, it is easy to misinterpret observed subtle differences in the SFRs of cluster and field galaxies as being due to environment, but which may be simply due to mis-matches in the stellar mass or redshift distributions of the two samples, or even differences in their sensitivity limits. For this reason we use field galaxies taken from exactly the same {\em Spitzer} images used to detect obscured star formation among the cluster galaxies. 

The use of the $J{-}K$ color selection to identify probable cluster galaxies to be targeted for spectroscopy in ACReS effectively produces a stellar mass-limited sample within a narrow redshift slice, which includes not only the cluster itself, but also field regions immediately in front of and behind it. 
For each cluster, we carefully determined the narrow redshift slices either side of the cluster for which field galaxies should still lie comfortably within the $J{-}K$ color slice. As our analysis of cluster galaxies is focused on those brighter than M$_{K}^{*}{+}1.5$, we also limited our field redshift slices ranges to those for which ACReS is complete to M$^{*}_{K}{+}1.5$. Given that the properties of galaxies in the infall regions of clusters are known to be systematically different from the field even at 3--4\,$r_{vir}$ \citep{chung11,wetzel11}, we exclude all galaxies within 4\,000\,km\,s$^{-1}$ of the cluster. We also remove galaxies at the same redshift as any X-ray group/cluster identified from our {\em XMM} map of the same region. Further details are given in the Appendix. 

The 30 clusters are evenly spaced through the redshift range of the survey, allowing us to build up a full and relatively even coverage of field galaxies over $0.15{<}z{<}0.30$ as the combination of many individual redshift slices perhaps just ${\sim}0$.02--0.05 wide. As each cluster is covered by {\em Spitzer} data of the same depths and fields of view, ultimately this field sample will have the same overall selection limits as the cluster counterpart.

\subsection{Bolometric infrared luminosities and SFRs}
\label{sec:lir}

For each of the 24$\mu$m-detected galaxies with known redshift, we estimated their intrinsic bolometric infrared luminosities \citep[$L_{TIR}$; as defined by][]{sanders} and rest-frame 24$\mu$m luminosities by comparison to template infrared spectral energy distributions (SEDs) obtained from observations of local star-forming galaxies. \citet{rieke} have constructed 14 average SED templates as a function of $L_{TIR}$. Each of these templates was shifted to the redshift and luminosity distance of the galaxy and convolved with the MIPS 24$\mu$m response function to obtain the predicted 24$\mu$m flux as a function of $L_{TIR}$. Our observed 24$\mu$m flux is compared to each of these values, and by interpolating between the templates, the intrinsic bolometric infrared and rest-frame 24$\mu$m luminosities estimated. The SFR is then estimated using the calibration of \citet{rieke}
\begin{equation}
{\rm SFR(M}_{\odot}{\rm yr}^{-1})= 7.8{\times}10^{-10}L(24{\mu}{\rm m}, L_{\odot}),
\label{sfrcalibration}
\end{equation}
which is valid for either a \citet{kroupa} or \citet{chabrier} IMF. In the case of cluster members, we use the luminosity distance corresponding to the cluster rather than the galaxy. Specific-SFRs are obtained by dividing this 24$\mic$-based SFR by stellar masses ($\mathcal{M}$) derived assuming a fixed $K$-band stellar mass-to-light ratio of 0.5, which should be appropriate for star-forming galaxies \citep{belldejong}.

\begin{figure}
\centerline{\includegraphics[width=84mm]{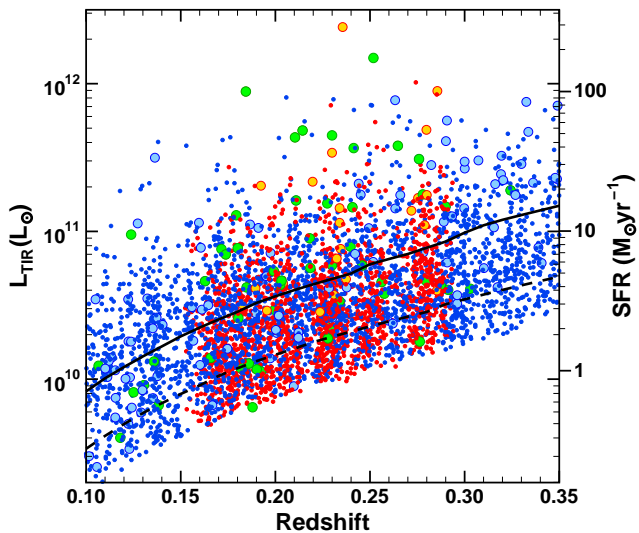}}
\caption{Total infrared luminosities (left-hand axis) and SFRs (right-hand axis) of the MIPS 24$\mu$m sources with a spectroscopic redshift at $0.1{<}z{\le}0.35$. 
Red symbols indicate cluster galaxies, blue symbols indicate field (i.e. non-cluster) galaxies. Larger orange/light blue symbols indicate cluster/field galaxies spectroscopically classified as Type I AGN from their broad emission lines. Large green symbols indicate X-ray point sources. The solid and dashed curves indicate as a function of redshift the infrared luminosity corresponding to observed 24$\mu$m fluxes of 1\,mJy and 0.4\,mJy (90\% completeness limit) respectively.} 
\label{lirz}
\end{figure}

 The resulting infrared luminosities and SFRs for 24$\mic$-detected galaxies in the redshift range $0.1{<}z{\le}0.35$ are shown in Fig.~\ref{lirz}, color coded as cluster members ({\em red}) and field galaxies ({\em blue}). 
At our survey 90\% completeness limit of 0.4\,mJy ({\em dashed line}), we are sensitive to $L_{TIR}{\sim}1{\times}10^{10}L_{\odot}$ for our low redshift ($0.15{<}z{\le}0.20$) clusters, rising to $L_{TIR}{\sim}3{\times}10^{10}L_{\odot}$ for our highest redshift ($z{\sim}0.28$) clusters. Our strategy of prioritizing for spectroscopy those objects with $f_{24}{>}1$\,mJy ({\em solid line}) should ensure that we are essentially complete for LIRGs ($L_{TIR}{>}10^{11}L_{\odot}$; ${\rm SFRs}{>}1$0\,M$_{\odot}$\,yr$^{-1}$) out to $z{=}0.30$. Most of the 24$\mu$m sources at $0.1{<}z{\le}0.35$ have $10^{10}{\la}L_{TIR}{\la}2{\times}10^{11}L_{\odot}$, although we identify also a few ULIRGs with $10^{12}{<}L_{TIR}{\la}2.5{\times}10^{12}L_{\odot}$. 

Among the sources with the highest infrared luminosities we see an increasingly high fraction that are likely powered by active nuclei, either identified by their X-ray emission ({\em green symbols}) or broad emission lines in their optical spectra (Type I AGN; {\em large orange/light blue symbols}). For cluster galaxies the AGN fraction among infrared-bright sources rises from 3\% at $L_{TIR}{<}10^{11}L_{\odot}$ to 65\% at $L_{TIR}{>}10^{11.6}L_{\odot}$, demonstrating the importance of being able to robustly account for the contribution of infrared emission powered by AGN, particularly when measuring the global SFRs of cluster members.

\subsection{Weighting to account for spectroscopic incompleteness}
\label{sec:weight}

To obtain robust measures of quantities such as the global cluster SFRs or the fraction of star-forming galaxies, we need to be able to account for incompleteness as a function of both location and galaxy property. This is done by weighting each galaxy by the inverse probability of it having been observed spectroscopically, following the approach of \citet{norberg}.

We give initial equal weight (1.0) to all those galaxies which could have been targeted for spectroscopy as described in {\S}~\ref{spectroscopy}, that is all galaxies with $J{-}K$ colors consistent with being a possible cluster member or having $f_{24}{>}1$\,mJy. For each galaxy lacking a redshift, its weight is transferred equally to its ten nearest neighboring galaxies with known redshift that also had the same priority level in our spectroscopic targeting strategy. This results in galaxies without redshifts having zero weight, while galaxies in regions where the spectroscopic survey is locally 50\% complete having weights of 2.0. The transferring of weight only within priority levels ensures that we can account statistically for the systematic differences in spectroscopic completeness from one level to another. The transferring of weight among neighboring galaxies, means that we are able to map the ``local'' variations in completeness.

\section{The mid-infrared luminosity function}
\label{sec:lf}

Many of the processes that drive the evolution of galaxies also shape the luminosity function (LF), one of the most basic and fundamental properties of the galaxy population \citep{benson}. The mid-infrared luminosity function can be used to measure the distribution of SFRs among a population of galaxies.  

\begin{figure}
\centerline{\includegraphics[width=84mm]{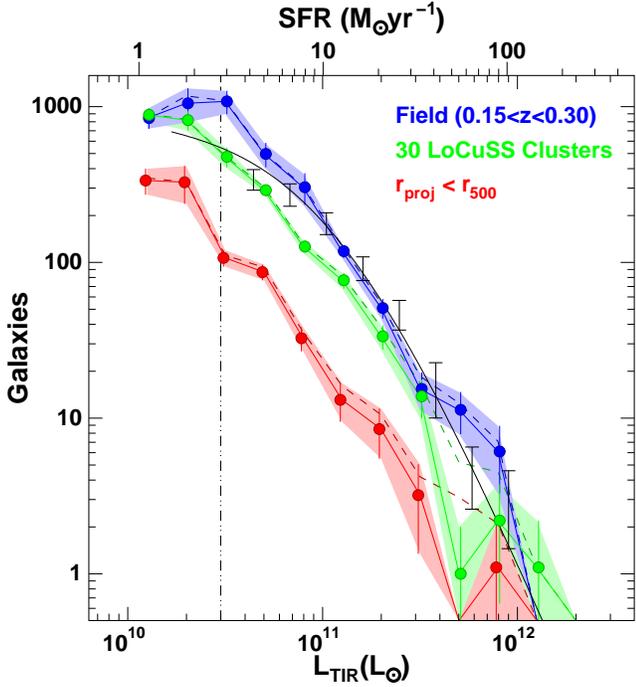}}
\caption{Stacked bolometric infrared luminosity function of the 30 LoCuSS clusters (green curve), including all cluster members with {\em Spitzer}/MIPS detections. The red curve includes just those cluster galaxies within $r_{500}$. The solid (dashed) curves indicate the LFs obtained by excluding (including) sources identified as X-ray AGN or QSOs. The blue curve shows the corresponding infrared luminosity function of spectroscopic field galaxies within the LoCuSS fields over the redshift range $0.15{<}z{<}0.30$. The vertical scale indicates the number of galaxies in each luminosity bin. The black curve and error bars indicate the infrared (24$\mu$m) luminosity function of star-forming field galaxies $0.20{<}z{<}0.35$ in the AGES survey from Figure 7 of \citet{rujopakarn}. The 24$\mu$m luminosities are converted to SFRs using our Eq.~\ref{sfrcalibration}.}
\label{lf}
\end{figure}

 We calculate the stacked infrared luminosity function for cluster galaxies from all 30 systems, including all cluster members covered by our {\em Spitzer} data and detected at 24$\mu$m. 
We bin galaxies according to their bolometric infrared luminosity ($L_{TIR}$). 
To account for the fact that we are sensitive to lower SFRs in cluster galaxies in the lowest redshift systems which would not be detected if placed in one of our clusters at higher redshift, we weight each galaxy by the inverse number of clusters for which it would have been detected in our {\em Spitzer} data given its measured 24$\mu$m luminosity. The resultant infrared luminosity function is shown by the solid green points in Fig.~\ref{lf}, and covers a factor 160 in luminosity, extending down to $L_{TIR}{=}10^{10}L_{\odot}$. Here we have excluded all those objects identified as X-ray AGN or spectroscopically classified as quasars, as in these cases the 24$\mu$m emission may well be dominated by dust heated by the AGN rather than star formation. Here, and throughout this paper, BCGs (defined as the brightest cluster galaxies in the cluster core) have been explicitly excluded due to their unique star formation histories and evolutions \citep{lin}, and the direct link between BCG activity and the presence of cooling flows within clusters \citep{edge,smith10a,rawle}.

At these redshifts ($0.15{<}z{<}0.30$), star-forming galaxies with $L_{TIR}{\ga}4{\times}10^{11}L_{\odot}$ (SFRs${\ga}4$0\,M$_{\odot}$\,yr$^{-1}$) appear to be extremely rare in cluster environments, with only four such galaxies, including one ULIRG, identified as cluster members among the full sample of 30 systems. The ULIRG ($L_{TIR}{=}10^{12.01}L_{\odot}$) is an ${\sim}L_{K}^{*}$ galaxy at $z{=}0.274$ located at a cluster-centric radius of 1.9\,Mpc (${\sim}1.0\,r_{200}$) from Abell 697, and based upon the Subaru imaging and its apparent extended 24$\mu$m emission it appears to be a merger. The optical spectrum, far-infrared SED 
and lack of detectable X-ray emission in the {\em Chandra} data, all support the view that the infrared emission is powered primarily by star formation rather than an AGN. 
As far as we are aware, this is the lowest redshift ULIRG identified as cluster member, BCGs excepted. 
Two BCGs are also classified as ULIRGs, those for the clusters Abell 1835 and ZwCl\,0857.9+2107 \citep{rawle}, and are the two most infrared-luminous sources reported in Fig.~\ref{lirz}. 

As the green points include cluster galaxies over the entire {\em Spitzer} fields of view, extending beyond the virial radius and into the infall regions, this will include many galaxies yet to encounter the cluster itself. We thus show as the solid red points and curve the infrared luminosity function for just those cluster members within a projected cluster-centric radius of $r_{500}$. 
The dashed green and red curves show the corresponding infrared luminosity functions, including now also those sources identified as AGN. This confirms what was apparent in Fig.~\ref{lirz}, that at the highest infrared luminosities ($L_{TIR}{\ga}3{\times}10^{11}L_{\odot}$) the contribution from AGN becomes dominant. 

The blue curve and points shows the infrared LF for star-forming field galaxies at $0.15{<}z{<}0.30$ that lie within the well-defined redshift limits for extracting field galaxies either side of each cluster as described in {\S}~\ref{sec:field} and shown in Fig~\ref{z_mk}. We find no ULIRGs among our field galaxy sample. Moreover, we do not find any ULIRGs at $z{<}0.35$ in any of the 30 {\em Spitzer} images, except those three identified above as cluster members (Fig~\ref{lirz}), confirming their rarity at these low redshifts. 
Our field galaxy infrared LF is consistent with that obtained by \citet[{\em black curve}]{rujopakarn} for $0.20{<}z{<}0.35$ star-forming field galaxies. 
The shape of the field galaxy infrared LF also appears fully consistent with that for our cluster galaxy sample. 
When limiting the comparison to luminosities where we should be highly complete at all redshifts ($L_{TIR}{>}3{\times}10^{10}L_{\odot}$; {\em dot-dashed line}), a two-sample Kolmogorov-Smirnov (K-S) test is unable to find significant differences between the $L_{TIR}$ distributions of star-forming galaxies in the two cluster samples and the field sample. 
This invariance of the infrared luminosity distribution of star-forming galaxies with environment is seen from the present day back to $z{\sim}0.8$, with the infrared LFs of galaxies in the Coma cluster and Shapley supercluster both found to be consistent with those of local field galaxies \citep{bai06,haines11a}, while \citet{finn10} found the infrared LFs of sixteen clusters at $0.42{<}z{<}0.80$ to be consistent with those of coeval field galaxies. 

\begin{figure}
\centerline{\includegraphics[width=84mm]{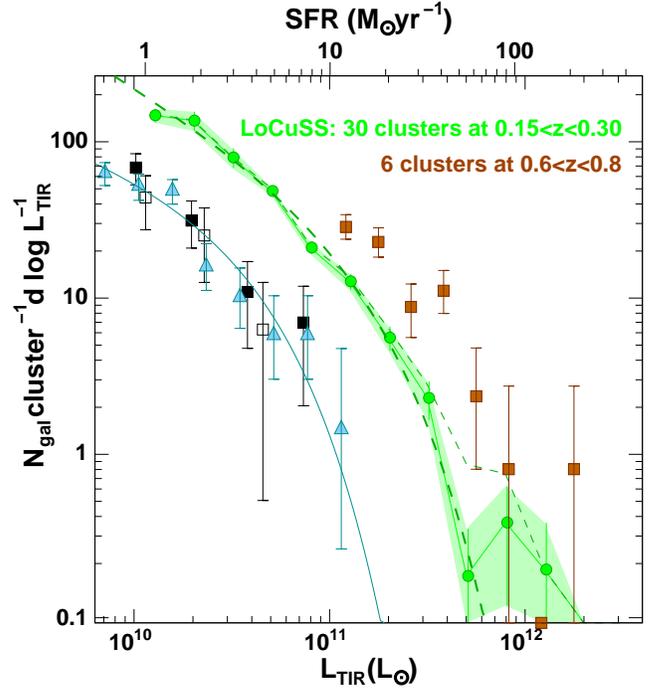}}
\caption{Evolution of the infrared luminosity function of cluster galaxies. The stacked infrared luminosity function of member galaxies from the 30 $0.15{\le}z{<}0.30$ clusters in our sample, excluding BCGs, X-ray AGN and QSOs (green points), reproduced from Figure~\ref{lf}, compared directly with the infrared luminosity functions of nearby systems -- Coma cluster \citep[$z{=}0.023$, open squares;][]{bai06}, Abell 3266 \citep[$z{=}0.06$, black squares;][]{bai09}, and the Shapley supercluster \citep[$z{=}0.048$, light blue triangles;][]{haines11a}; and the composite infrared LF of six rich clusters at $0.6{<}z{<}0.8$ (brown squares) from \citet{finn10}. To ease comparison between epochs we have normalized the vertical scale as the number of galaxies per cluster per unit dex in $L_{IR}$. The cyan curve shows the best-fit Schechter function to the infrared LF of the Shapley supercluster \citep[$L_{IR}^{*}{=}10^{10.52}L_{\odot}$, $\alpha{=}{-}1.425$;][]{haines11a}. The green dashed curve shows the best-fit Schechter function for our 30 systems, with $L_{IR}^{*}{=}10^{11.11}L_{\odot}$; $\alpha{=}{-}1.75$.
}   
\label{lf_evol}
\end{figure}

\subsection{Evolution of the mid-infrared LF since $z{\sim}0.8$}
\label{sec:lf_evol}

We chart the evolution of the mid-infrared luminosity function of cluster galaxies from $z{\sim}0.8$ to the present day in Figure~\ref{lf_evol}, comparing the stacked infrared LF of  the 30 $0.15{\le}z{<}0.30$ clusters in our sample ({\em green points}) derived in Section~\ref{sec:lf}, with the previously published infrared LFs of galaxies from the Coma cluster at $z{=}0.023$, Abell 3266 at $z{=}0.06$ and the Shapley supercluster at $z{=}0.048$, as well as the composite infrared LF of six rich clusters at $0.6{<}z{<}0.8$ from \citet{finn10}. 

Firstly we find that the three published infrared LFs of the local ($z{\le}0.06$) systems are fully consistent with one another, and can be well described by the best-fit Schechter function ({\em cyan curve}) to the infrared LF of the Shapley supercluster \citep{haines11a}, providing a robust and established base-line for star formation in present day rich clusters. When comparing this $z{\sim}0$ base-line LF to the infrared LF averaged over the 30 clusters in our sample, clear evolution is apparent even at these modest redshifts. While we note that the systems from our LoCuSS sample are slightly richer on average than those from our $z{\sim}0$ cluster sample, shifting the normalized infrared LF upwards, the bulk of the difference between the two LFs is likely due to a factor ${\sim}3{\times}$ increase in the characteristic infrared luminosity $L_{IR}^{*}$ of cluster galaxies. It is simply not possible to explain the significant numbers of galaxies with $L_{IR}{\ga}2{\times}10^{11}L_{\odot}$ in our $0.15{\le}z{<}0.30$ clusters by only shifting the local infrared LF upwards in the plot (density evolution) given the complete absence of such galaxies in local clusters. 
In {\S}~\ref{sec:ssfr} we shall see that the specific-SFRs of cluster galaxies have declined by a factor ${\sim}3{\times}$ since $z{\sim}0.3$ (Fig.~\ref{ssfr_evol}), which would naturally explain a ${\sim}3{\times}$ decline in $L_{IR}^{*}$ and the observed evolution in the mid-IR LFs over the same period.

\citet{finn10} recently presented the composite infrared LFs of six rich clusters at $0.6{<}z{<}0.8$ from the ESO Distant Cluster Survey allowing us to follow the cluster infrared LF beyond $z{=}0.3$. Although their lower luminosity limit of $10^{11}L_{\odot}$ makes it impossible to fully distinguish between luminosity and density evolution scenarios, a qualitative comparison between the three infrared LFs is consistent with a simple scenario in which the infrared luminosity function of cluster galaxies has evolved from $z{\sim}0.8$ to the present day via a steady and continual decline in $L_{IR}^{*}$ amounting to a factor ${\sim}6{\times}$ drop over the intervening period.

\begin{figure}
\centerline{\includegraphics[width=84mm]{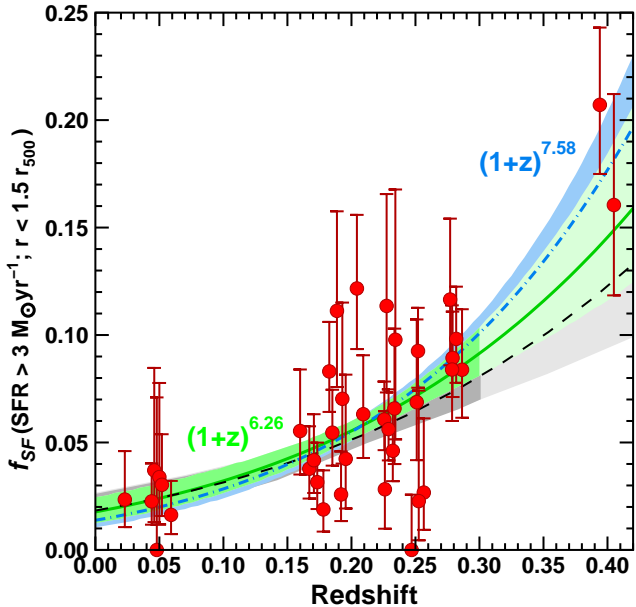}}
\caption{The mid-infrared Butcher-Oemler effect.  The red symbols show the star-forming fraction $f_{SF}$ for each cluster, defined as the estimated fraction of $M_{K}{<}{-}23.10$ ($M_{K}^{*}{+}1.5$) cluster members within 1.5\,r$_{500}$ having ${\rm SFR}{>}3.0\,{\rm M}_{\odot}$\,yr$^{-1}$ ($L_{TIR}{>}3{\times}10^{10}L_{\odot}$). The error bars indicate the uncertainties derived from binomial statistics calculated using the formulae of \citet{gehrels}. 
The green curve indicates the best-fitting evolutionary curve of the form $f_{SF}{=}f_{0}\,(1+z)^{n}$, for the 37 clusters at $z{<}0.3$. The shaded region indicates the 1$\sigma$ confidence region to that fit, with the lighter colors showing the extrapolation of the fit beyond $z{=}0.3$. The blue dot-dashed curve and shaded region indicates the same fit, including also the two clusters at $z{\sim}0.4$.  
The black dashed line indicates the corresponding evolutionary curve from the previous analysis of \citet{haines09a}. The shaded region indicates the 1$\sigma$ confidence region to that fit, with the lighter colors showing the extrapolation of the fit beyond $z{=}0.3$.} 
\label{bo}
\end{figure}

\section{The Mid-Infrared Butcher-Oemler effect}
\label{sec:results}
\label{sec:bo}

For each cluster, we compute the fraction of star-forming cluster members ($f_{SF}$) within a projected cluster-centric radius of $1.5\,r_{500}$ \citep[${\sim}1.0\,r_{200}$;][]{sanderson03}. This is defined as the fraction of $K$-band selected galaxies with $M_{K}{<}{-}23.10$ ($M_{K}{<}M_{K}^{*}{+}1.5$; Vega) having ${\rm SFRs}{>}3.0\,{\rm M}_{\odot}{\rm yr}^{-1}$ 
($L_{TIR}{>}3{\times}10^{10}\,L_{\odot}$). 
To derive the fractions, we use only those $M_{K}{<}M_{K}^{*}{+}1.5$ galaxies spectroscopically identified as cluster members, weighted to account for incompleteness as described in {\S}~\ref{sec:weight}. Uncertainties in $f_{SF}$ are derived from binomial statistics using the formulae of \citet{gehrels}.

In Figure~\ref{bo} we chart the evolution of $f_{SF}$ with redshift -- the mid-infrared Butcher-Oemler effect -- for 39 clusters over $0.0{<}z{<}0.40$. This shows a steady increase in $f_{SF}$ with redshift from $\langle f_{SF}\rangle{=}0.023{\pm}0.013$ at $z{\le}0.06$, through $\langle f_{SF}\rangle{=}0.052{\pm}0.026$ at $0.15{<}z{<}0.20$ to $\langle f_{SF}\rangle{=}0.073{\pm}0.033$ at $0.20{<}z{<}0.30$, and if we include the two highest redshift non-LoCuSS clusters $f_{SF}{=}0.185{\pm}0.025$ at $z{\sim}0.40$. 

To quantify the redshift evolution of $f_{SF}$, we fit the relation $f_{SF}{=}f_{0}(1{+}z)^{n}$ to the 37 $z{<}0.3$ clusters shown in Fig.~\ref{bo}. 
The best-fit relation is shown by the green curve, and has an exponent $n{=}6.26_{-1.54}^{+1.70}$, consistent with the value $n{=}5.7_{-1.8}^{+2.1}$ obtained in Paper I. 
The new fit is marginally higher than that from Paper I ({\em black dashed curve}) at all redshifts (at the ${\sim}1{\sigma}$ level), which is expected given the slightly lower SFR limit used here of 3\,M$_{\odot}$\,yr$^{-1}$ rather than the 5\,M$_{\odot}$\,yr$^{-1}$ limit used in Paper I. 
If we include also the two clusters at $z{\sim}0.4$, the relation steepens slightly to $n{=}7.58_{-1.14}^{+1.16}$ ({\em blue dot-dashed curve}).

\begin{figure}
\centerline{\includegraphics[width=84mm]{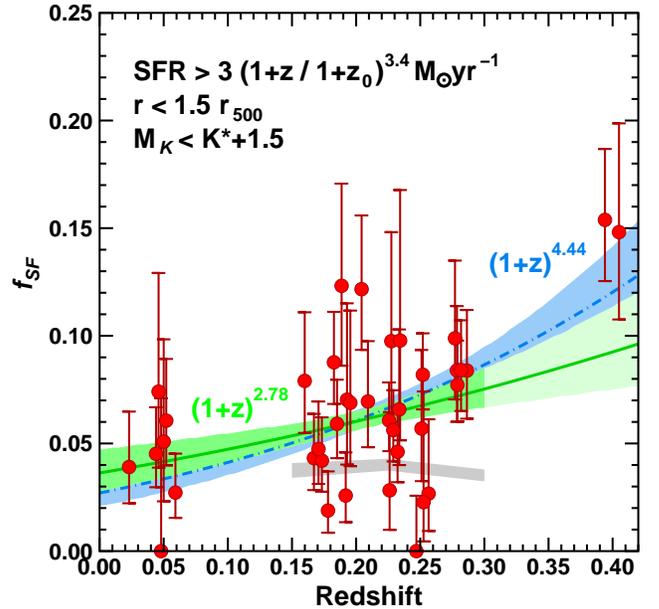}}
\caption{As in Figure~\ref{bo}, but using an evolving SFR threshold that matches the observed cosmic decline in the SFRs of field galaxies of the form $(1+z)^{3.4}$ \citep{rujopakarn}. 
The green curve indicates the best-fitting evolutionary curve of the form $f_{SF}{=}f_{0}\,(1+z)^{n}$, for the 37 clusters at $z{<}0.3$. The shaded region indicates the 1$\sigma$ confidence region to that fit, with the lighter colors showing the extrapolation of the fit beyond $z{=}0.3$. The blue dot-dashed curve and shaded region indicates the same fit, including also the two clusters at $z{\sim}0.4$.  
The gray shaded region indicates the estimated contribution to $f_{SF}$ from interlopers which are actually at $r{>}r_{200}$.} 
\label{boevol}
\end{figure}

\subsection{The mid-IR Butcher-Oemler effect in the context of the cosmic decline in star formation}
\label{sec:boevol}

In Paper I we argued that the Butcher-Oemler effect could simply be due to the cosmic decline in star formation among field galaxies which are subsequently accreted onto the clusters. We noted that while the characteristic infrared luminosity $L_{TIR}^{*}$ (or equivalently SFR) of both cluster and field galaxies has declined rapidly since $z{\sim}1$ \citep{lefloch,rujopakarn}, our SFR threshold had been kept constant, and hence was not selecting the same kinds of star forming galaxies at $z{\sim}0.4$ as at the present day, resulting in a decline in $f_{SF}$ with time. To counteract this, the SFR threshold should instead be kept fixed with respect to $L_{TIR}^{*}(z)$. We found that when doing this, the bulk of the redshift evolution in $f_{SF}$ disappeared, suggesting that the primary cause of the Butcher-Oemler effect was the decline in star formation among those field galaxies being accreted onto the cluster, rather than a cluster-specific phenomenon. There did however, remain a residual evolution among cluster galaxies of the form $f_{SF}{\propto}(1+z)^{1.2{\pm}1.4}$, after accounting for the cosmic decline in $L_{TIR}^{*}$, which appeared mostly confined to $1.0{\la}(r/r_{500}){\la}1.5$. Indeed we were able to obtain a good fit to the data with a fixed $f_{SF}$ when considering only galaxies within $r_{500}$. 

Following Paper I, we replace the fixed SFR limit used in Fig.~\ref{bo} with an evolving SFR threshold of the form ${\rm SFR}>3.0\,((1+z)/(1+z_{0}))^{3.4}$\,M$_{\odot}$\,yr$^{-1}$ ($z_{0}{=}0.225$) to match the observed cosmic decline in $L_{TIR}^{*}(z){\propto}(1+z)^{3.4}$ obtained by \citet{rujopakarn}, to produce Figure~\ref{boevol}. 
Unlike in Paper I, there remains a significant trend with $f_{SF}$ increasing with redshift as $(1+z)^{2.78{\pm}1.33}$ ({\em green curve}). The reduction in the $f_{SF}$ of cluster galaxies since $z{\sim}0.3$ is thus inconsistent with being simply due to the cosmic decline in star formation at the 2.1$\sigma$ significance level. The residual evolution is even more significant if the two $z{\sim}0.4$ clusters are also included, with a best-fit relation of the form $f_{SF}{\propto}(1+z)^{4.44{\pm}1.21}$ ({\em blue dot-dashed curve}). 
The mid-infrared Butcher-Oemler effect requires additional environmental processes acting to reduce the SFRs of cluster galaxies since $z{\sim}0.4$. 
We shall show in {\S}~\ref{sec:ssfr} that the specific-SFRs of cluster galaxies were declining faster than their counterparts in the field over this period.

\begin{figure}
\centerline{\includegraphics[width=84mm]{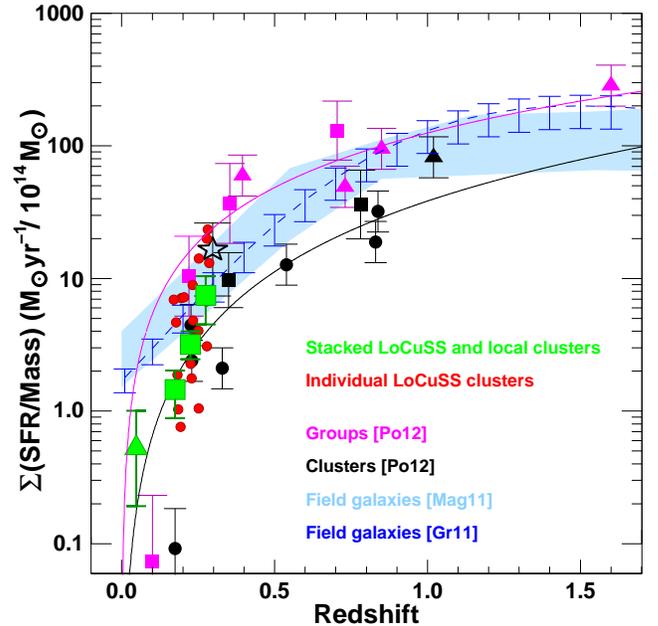}}
\caption{Evolution of the total SFRs among cluster LIRGs per unit halo mass. The red points indicate the total star formation rate for all the cluster LIRGs within $r_{200}$ for each LoCuSS cluster normalized by the cluster mass ($M_{200}$ in units of $10^{14}{\rm M}_{\odot}$) as a function of redshift. The solid green squares indicate the averaged values for the LoCuSS clusters in three redshift bins, combining 11 systems at $0.15{<}z{<}0.20$, 10 systems at $0.20{<}z{<}0.25$ and 9 systems at $0.25{<}z{<}0.30$. The green triangle shows the composite local systems averaged over the five clusters in the Shapley supercluster, plus Coma and Abell 3266. The error bars indicate the bootstrap uncertainties based on randomly sampling N clusters with replacement from the N clusters in each bin. The black and magenta points show the $\Sigma$(SFR)/$M$ values for clusters and groups respectively from \citet{popesso12} (Po12; their Figure 3), after normalizing their SFRs down by a factor 1.7 to match our IMF. Square symbols, triangles and dots identify respectively, their COSMOS composite systems, the GOODS composite systems and other individual systems. The star indicates the Bullet cluster. The black solid line indicates the best-fit $\Sigma$(SFR)/{\em M}--{\em z} relation for their cluster sample, excluding the Bullet cluster.  The magenta curve indicates the corresponding relation for their groups and poor clusters. Both relations are of the form $\Sigma{\rm (SFR)/M}{\propto}z^{\alpha}$. The field $\Sigma$(SFR)/{\em M}--{\em z} relation from \citet{magnelli} (Mag11; light blue shaded region) and \citet{gruppioni} (Gr11; dashed blue line) are also shown. The shading and error bars represent the 1$\sigma$ confidence levels. 
}
\label{popesso}
\end{figure}

The steepening of the evolutionary trend with respect to that of Paper I appears primarily due to the shift from having to rely on the statistical subtraction of field galaxies from control fields to estimate the $f_{SF}$ in Paper I, to being able to estimate the $f_{SF}$ solely from spectroscopic cluster members. 

\subsection{Star formation activity per unit halo mass}

A frequently used metric of the evolution of star formation activity among cluster galaxies is to consider the globally measured star formation rate per unit cluster mass, that is the sum of the SFRs of all the confirmed cluster members within $r_{200}$ divided by the total mass of the system ($M_{200}$), $\Sigma$(SFR)/$M$. This allows systems of widely differing masses to be easily combined or compared, and can also be considered a way of quantifying the efficiency of forming stars and building up stellar mass as a function of halo mass. 

In the previous section we saw a rapid decline in the level of star formation activity in cluster galaxies since $z{\sim}0.4$. 
To understand this decline in the wider context of cluster galaxy evolution over the last 10 billion years, we combined our data with the analysis of \citet[hereafter Po12]{popesso12} to produce Figure~\ref{popesso}. This reproduces their Figure~3, in which each point indicates the total SFRs of luminous infrared galaxies (LIRGs) within $r_{200}$ normalized by halo mass, for both rich clusters ($M_{200}{\sim}10^{15}M_{\odot}$; {\em black symbols}) and groups/poor clusters ($M_{200}{\sim}10^{13.5}M_{\odot}$; {\em magenta symbols}) over $z{\sim}0$.1--1.6 based on {\em Herschel}/PACS 100$\mu$m and 160$\mu$m imaging. 

\begin{figure*}
\centerline{\includegraphics[width=130mm]{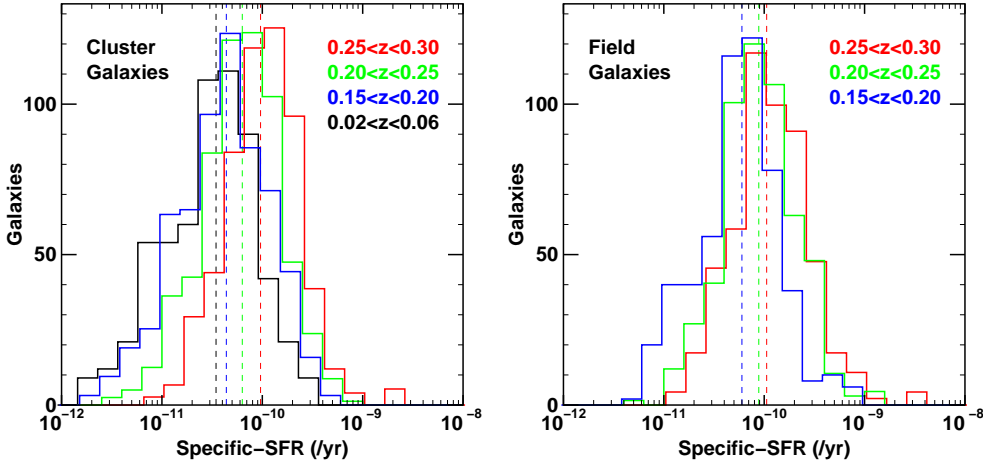}}
\caption{The evolution of the specific-SFR distribution of cluster ({\em left panel}) and field ({\em right panel}) star-forming galaxies. The histograms show the specific-SFRs of $M_{K}{<}-23.1$ galaxies detected in the LoCuSS {\em Spitzer}/MIPS 24$\mu$m images for three redshift bins: $0.15{<}z{<}0.20$ (blue); $0.20{<}z{<}0.25$ (green); $0.25{<}z{<}0.30$ (red). The black histogram in the left panel shows the specific-SFR distribution of star-forming galaxies in the Shapley supercluster at $z{=}0.048$. Vertical dashed lines indicate the biweight estimator of the mean ($C_{BI}$) of each distribution.}
\label{ssfr_evol}
\end{figure*}

Following Po12 we summed up the SFRs of all the cluster LIRGs (excluding X-ray AGN, BCGs and QSOs) found within $r_{200}$ for each of the 30 clusters in our sample, and divided the total by the cluster mass, $M_{200}$, derived from our {\em Chandra} X-ray data \citep{sanderson09}. For the 21 out of 30 clusters which contain at least one LIRG, the resulting $\Sigma$(SFR)/$M$ values are indicated by red points. 
To measure the overall evolutionary trend, the clusters are split into three redshift bins containing 11 systems at $0.15{<}z{<}0.20$, 10 systems at $0.20{<}z{<}0.25$ and 9 systems at $0.25{<}z{<}0.30$, which are shown by the green squares. The error bars indicate the uncertainties due to the cluster-to-cluster scatter in $\Sigma$(SFR)/$M$, estimated by bootstrap resampling the clusters in each bin. 
Even within the narrow redshift slice covered by LoCuSS, corresponding to just 1.5\,Gyr in look-back time, we observe a factor ${\sim}5{\times}$ decline in the level of star formation activity per unit halo mass. This rapid decline reflects both a reduction in the numbers of cluster LIRGs, from 25 in the nine $0.25{<}z{<}0.30$ clusters to just 8 in the 11 systems at $0.15{<}z{<}0.20$, but also a reduction in the SFRs in cluster LIRGs on average. This can be seen from Figure~\ref{lirz}, which reveals that for the lowest redshift systems, essentially all the cluster LIRGs have $L_{IR}{\la}2{\times}10^{11}L_{\odot}$, while at higher redshifts there are increasing numbers of LIRGs with $L_{IR}{\ga}2{\times}10^{11}L_{\odot}$ (SFRs${\ga}2$0\,M$_{\odot}$yr$^{-1}$). 

To see if this decline extends right to the present day, the infrared data for the Coma cluster, Abell 3266 and the five systems from the Shapley supercluster are combined. We find just two LIRGs within $r_{200}$ for these seven clusters, one in Abell 3266 \citep{bai09} and one in Shapley \citep{haines11a}, resulting in an even lower value of $\Sigma$(SFR)/$M$ ({\em green triangle}) than seen in the lowest redshift LoCuSS systems. 
Overall we measure a decline of a factor ${\sim}15{\times}$ in $\Sigma$(SFR)/$M$ for cluster LIRGs over the last 3.3\,Gyr, since $z{=}0.3$. This is roughly consistent with the best-fit evolutionary trend of Po12 ($\Sigma$(SFR)/$M{\propto}\,z^{1.77{\pm}0.36}$; {\em black curve}), and indicates that the decline in star formation activity among cluster galaxies has become much more rapid since $z{\sim}0.3$ than over $z{\sim}0$.3--1.0. This is consistent with mid-IR Butcher--Oemler analysis of \citet{saintonge} who find that the trend for $f_{SF}$ remains essentially flat ($f_{SF}{\sim}0.13$) over $z{\sim}0$.4--0.83, albeit based on just five systems at these redshifts (Fig. 2 of Paper I). 

The ${\sim}15{\times}$ decline in $\Sigma$(SFR)/$M$ since $z{\sim}0.30$ seen for the cluster galaxies in our sample is much greater than the factor ${\sim}$3--5${\times}$ decline seen for field galaxies \citep[{\em blue curves};][]{gruppioni,magnelli} over the same period. It thus appears that the rapid reduction in star formation activity of cluster LIRGs since $z{\sim}0$.3--0.4 cannot be ascribed solely to the steady {\em cosmic} decline in star formation seen in field galaxies, but requires an additional suppressional impact from environmental processes acting within the clusters themselves. 

This extremely sharp drop off in $\Sigma$(SFR)/$M$ since $z{\sim}0.3$ is partly due to our limiting the analysis to cluster LIRGs, as can be understood from Figure~\ref{lf_evol}. While at high redshifts, the LIRG limit is well below $L_{IR}^{*}(z)$ and so the sample is dominated by normal star-forming galaxies, by $z{\sim}0$.2--0.3, our cut off at $L_{IR}{=}10^{11}L_{\odot}$ is rising through the knee of the luminosity function. Moving forward to the present day, as the limit now passes through the range in the luminosity function dominated by the exponential function, the number density of LIRGs drops exponentially with time, producing the sharp observed decline. We do note however that since the infrared luminosity function appears invariant with environment, this should affect the evolution in the number densities of both cluster and field LIRGs in the same manner, and so this cannot account for the more rapid decline seen in the star formation activity of cluster LIRGs.

\section{The decline in specific-SFRs among cluster and field galaxies since z${\sim}$0.3}
\label{sec:ssfr}

\begin{figure*}
\centerline{\includegraphics[width=170mm]{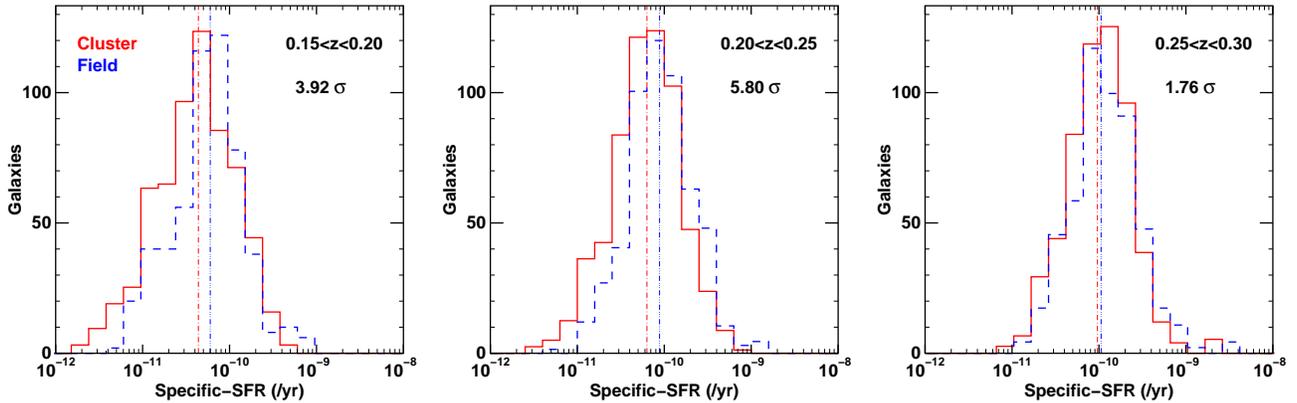}}
\caption{Comparison of the specific-SFR distributions of cluster (red histograms) and field (blue dashed histograms) galaxies detected in the LoCuSS {\em Spitzer}/MIPS 24$\mu$m images for three redshift bins: $0.15{<}z{<}0.20$ ({\em left panel}); $0.20{<}z{<}0.25$ ({\em centre}); $0.25{<}z{<}0.30$ ({\em right panel}). Vertical red/blue dot-dashed lines indicate the biweight estimator of the mean ($C_{BI}$) of each distribution. The significance of the differences between the cluster and field specific-SFR distributions for each redshift bin are indicated in units of $\sigma$.}
\label{ssfr_compare}
\end{figure*}

A key question to ask is whether the rapid decrease in star formation activity among cluster galaxies shown by Figures~\ref{bo} and~\ref{popesso} reflects: (i) a decline in the numbers or fraction of cluster galaxies with ongoing star formation, while leaving the SFRs of those star-forming galaxies invariant --- {\em density evolution}; or (ii) a systematic decline in the SFRs of star forming cluster galaxies while the numbers of those star-forming galaxies remain constant --- {\em luminosity evolution}.
The analysis of {\S}~\ref{sec:lf_evol} and Fig~\ref{lf_evol} suggests a significant decline in the SFRs of cluster galaxies since $z{\sim}0.8$, but given the relatively steep faint-end slopes found for the 24$\mu$m LFs for each redshift bin, there are no significant breaks in the LFs to robustly constrain $L_{IR}^{*}(z)$, making it impossible to fully separate the effects of luminosity and density evolution with any degree of confidence. 

A better approach is to consider instead the specific-SFR distribution of cluster galaxies as a function of redshift. In the field, the decline in star formation in galaxies isn't due to a change in the stellar mass function of star forming galaxies \citep[it is remaining constant;][]{bell}, but rather a systematic decline in the specific-SFRs of star forming galaxies at all masses \citep{noeske,elbaz}. We can thus equate the evolution in the specific-SFR distribution (at fixed stellar mass) with luminosity evolution. 
The particular advantage of quantifying luminosity evolution via the specific-SFR distribution, is that for mass-limited samples, the specific-SFR distribution is approximately log-normal in form \citep{noeske,bothwell}. Hence assuming that the survey is sensitive enough to extend below the central peak in specific-SFR (sSFR), one can robustly determine the value of $\langle{\rm sSFR}(z)\rangle$, unlike $L_{IR}^{*}(z)$ which can remain unconstrained. 

Figure~\ref{ssfr_evol} shows the specific-SFR distributions for cluster ({\em left panel}) and field galaxies ({\em right panel}) in the LoCuSS {\em Spitzer}/MIPS images for three redshift bins over $0.15{<}z{<}0.30$. Only $M_{K}{<}{-}23.10$ ($M_{K}^{*}{+}1.5$) galaxies detected at 24$\mic$ are included, while those identified as X-ray AGN or QSOs are removed. The biweight estimators of the mean \citep[$C_{BI}$;][]{beers} and standard deviation $S_{BI}$ for each distribution of log(sSFR) are tabulated in Table~\ref{table:ssfr}. 

 In all redshift bins and environments the specific-SFR distribution of massive galaxies shows an approximately log-normal form, with a single, central peak and a relatively narrow distribution ($\sigma{\sim}0.4$\,dex). There is some asymmetry, with a heavy tail at low specific-SFRs, particularly at low redshifts. 

There is clear evolution in the specific-SFRs of galaxies, both in clusters and in the field. 
The specific-SFRs of all star-forming galaxies appear to have declined by a factor 1.7--2.2 just within the narrow redshift slice (0.15--0.30) of the LoCuSS survey, both in cluster environments and in the field. The offsets in the specific-SFR distributions between the $0.25{<}z{<}0.30$ and $0.15{<}z{<}0.20$ redshift bins are significant at the 11.26$\sigma$ level for the cluster sub-sample and 7.26$\sigma$ level for the field sub-sample. 
If we consider also the data from the Shapley supercluster at $z{=}0.048$ ({\em black histogram} in left panel) the specific-SFRs of star-forming galaxies in clusters have declined by a factor $2.78{\pm}0.24$ since $z{\sim}0.30$. This decline in the specific-SFRs can be seen not only in the shifting leftward of the central peak of the distribution, but also the upper and lower tails. The {\em entire} main sequence of star-forming galaxies is shifting towards lower specific-SFRs with time, both in clusters and in the field. 

We note that some of this apparent evolution is due to selection bias. Within the LoCuSS sample, the same 24$\mic$ photometry is sensitive to SFRs and specific-SFRs a factor ${\sim}2$.5--3 lower for galaxies in the low-redshift ($0.15{<}z{<}0.20$) bin than in the high-redshift ($0.25{<}z{<}0.30$) bin. This selection bias affects the numbers of low specific-SFR galaxies (${\la}2{\times}10^{-11}\,{\rm yr}^{-1}$) in the highest redshift bin, as many such galaxies will fall below our 24$\mu$m completeness level. 
It cannot however explain the observed systematic shift downwards by 0.2--0.3\,dex of the central peak and upper end of the specific-SFR distribution from the high redshift to low-redshift bins.

\begin{table}
\centering
\begin{tabular}{ccccccc} \hline
Redshift & \multicolumn{3}{c}{Cluster} & \multicolumn{3}{c}{Field}\\
Range & $\langle{\rm sSFR}\rangle$&$\sigma$(sSFR)&$\langle{z}\rangle$& $\langle{\rm sSFR}\rangle$&$\sigma$(sSFR)&$\langle{z}\rangle$\\ \hline
0.03--0.06&-10.46&0.46&0.0480&---&---&---\\
0.15--0.20&-10.36&0.46&0.1791&-10.22&0.41&0.1778\\
0.20--0.25&-10.20&0.40&0.2222&-10.05&0.37&0.2242\\
0.25--0.30&-10.02&0.36&0.2691&-9.98&0.37&0.2769\\ \hline 
\end{tabular}
\caption{Mean and dispersion of specific-SFRs of cluster and field galaxies as a function of redshift.}
\label{table:ssfr}
\end{table}

\subsection{Comparison of the decline in specific-SFRs of cluster and field galaxies}

 The trends revealed by Figure~\ref{ssfr_evol} support the view that the decline in the specific-SFRs of cluster galaxies (luminosity evolution) plays an important part in the observed Butcher-Oemler effect. It also qualitatively supports the claim from Paper I that the decline in star formation among cluster galaxies since $z{\sim}0.4$ is primarily a consequence of the decline seen in star-forming field galaxies, which are subsequently accreted onto the clusters. We can separate the relative importance of this latter effect by comparing the specific-SFR distributions of cluster and field populations at fixed redshift. If the decline in the specific-SFR of cluster galaxies is solely due to that seen in field galaxies, their specific-SFRs distributions should be identical, or at least decline in step. If however, there is some {\em added} evolution due to cluster-specific processes we should (but not necessarily, e.g. if quenching is rapid) see some systematic differences in the specific-SFR distributions of cluster and field galaxies at fixed redshift, in the form of a more rapid decline in the specific-SFRs of cluster galaxies. 

In Figure~\ref{ssfr_compare} we directly compare the specific-SFR distributions of 24$\mic$-detected galaxies in cluster ({\em red solid histograms}) and field ({\em blue dashed histograms}) environments for each of the three narrow redshift slices covered by the LoCuSS survey ($0.15{<}z{<}0.20$; $0.20{<}z{<}0.25$; $0.25{<}z{<}0.30$). The significance levels of the differences between the cluster and field specific-SFR distributions for each redshift bin, as estimated using the Mann-Whitney U-test, are shown in each panel. There should be no selection bias between the cluster and field subsamples at fixed redshift, as all of our {\em Spitzer}/MIPS 24$\mic$ images have the same exposure times and sensitivities. In other words, a cluster galaxy of a given specific-SFR should be equally likely to be detected at 24$\mic$ and be targeted for spectroscopy on average as a field galaxy of the same specific-SFR and redshift (but located in one of the other LoCuSS 24$\mic$ images). 

For the $0.15{<}z{<}0.20$ and $0.20{<}z{<}0.25$ redshift bins, the specific-SFRs of star-forming galaxies in clusters are systematically ${\sim}0.15$\,dex (${\sim}3$0\%) lower than their counterparts in the field, offsets significant at the 3.92$\sigma$ and 5.80$\sigma$ levels respectively. The difference between the two distributions is much greater on the low specific-SFR tail (${\la}3{\times}10^{-11}{\rm yr}^{-1}$) where we see a clear excess of cluster galaxies, than on the high specific-SFR side. 
This difference is much less pronounced in the high redshift bin, although even here the mean specific-SFR of 24$\mic$-detected cluster galaxies is marginally (${\sim}0$.04\,dex) lower than their field counterparts. 
This reduced difference is again likely due to incompleteness at low specific-SFRs for the most distant cluster subsample.

Given the rapid evolution in the specific-SFRs of star-forming galaxies revealed by Fig.~\ref{ssfr_evol}, we may be concerned that the differences in the specific-SFRs of cluster and field galaxies could be due to the cluster galaxies being at slightly lower redshifts than their field counterparts, despite the narrow redshift bins used. However, we confirm that for the lower two redshift bins where significant differences in the specific-SFRs of cluster and field galaxies are seen, the mean redshifts of the two samples are within 0.002 of one another (Table~\ref{table:ssfr}).

There is no evidence of {\em triggered} star formation in the cluster environment, in the form of an excess number of cluster star-forming galaxies with high specific-SFRs (or starbursts) with respect to the field.

\subsection{Comparing the specific-SFRs of cluster and field galaxies at fixed stellar mass}
\label{sec:quenching}

 In Figure~\ref{ssfr_compare} we identified a significant population of star-forming galaxies in clusters with {\em reduced} specific-SFRs (${\la}3{\times}10^{-11}{\rm yr}^{-1}$), but which retained sufficient star formation to be detectable by our {\em Spitzer}/MIPS 24$\mic$ images. \citet{balogh04} argued that this systematic reduction in specific-SFRs (or equivalently the equivalent width of H$\alpha$ emission) of {\em star-forming} galaxies requires that a significant fraction of these star-forming galaxies must be {\em currently} having their star-formation quenched by some process initiated when they were accreted by the cluster. For this to be simultaneously occurring in so many cluster galaxies requires that the quenching time-scale to be long enough (${\ga}1$\,Gyr) to ensure sufficient {\em quenching} galaxies with {\em reduced, but detectable} star-formation to skew the specific-SFR distribution. 
If the quenching process were rapid (on ${\la}1$00\,Myr time-scales), the 24$\mic$ emission from star formation in these galaxies would rapidly fall below our sensitivity limit, leaving few if any cluster galaxies with reduced, but detectable 24$\mic$ emission (SFRs). This would leave the mean and $\sigma$ of the specific-SFR distribution of the remaining star-forming galaxies largely unaffected, although they would be fewer in number.

A major caveat of the previous analysis is that the specific-SFRs of galaxies on the main star-forming sequence depends on their stellar mass, being systematically lower for more massive galaxies \citep{salim,bothwell}. This decline of specific-SFRs with stellar mass has also been observed for star-forming cluster galaxies \citep{wolf,lu}. 
Hence, the reduced specific-SFRs of star-forming galaxies in clusters could be explained if they were more massive on average than star-forming galaxies in the field.

\begin{figure}
\centerline{\includegraphics[width=84mm]{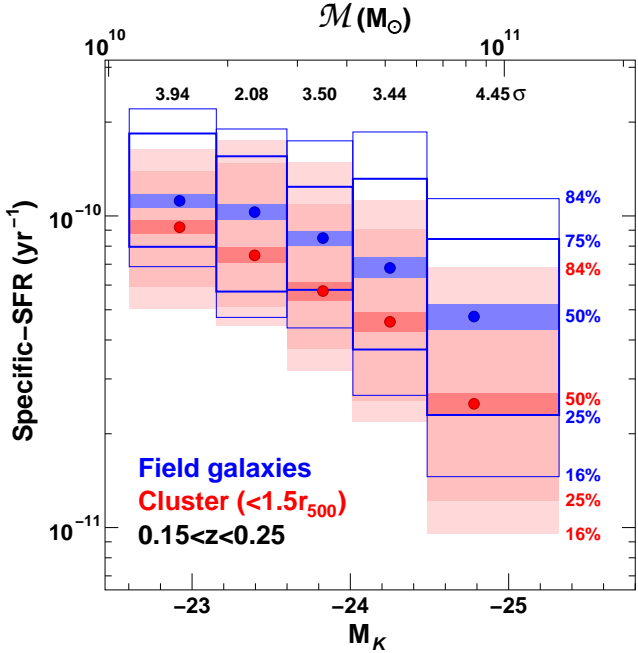}}
\caption{Comparison of the main sequence of star-forming galaxies in cluster ($r_{proj}{<}1.5\,r_{500}$) and field environments at $0.15{<}z{<}0.25$. The red (blue) circles show the median specific-SFRs of cluster (field) star-forming (detected at 24$\mic$) galaxies in bins of stellar mass. Each bin is defined to contain the same number (300) of galaxies combining both cluster and field subsamples. The vertical spread of the darkest shaded regions enclosing each circle indicate the uncertainty in this median value. The two further graduations in the red shaded regions indicate the interquartile and 1$\sigma$ (16--84\%) ranges of specific-SFRs for cluster galaxies in each stellar mass bin. The two blue boxes indicate the same ranges in specific-SFRs for the star-forming field galaxies. The values at the top of each mass bin indicate the significance (in $\sigma$) of the differences between the observed cluster and field specific-SFR distributions for that stellar mass bin.}
\label{ssfr_kmag}
\end{figure}

 To address this, the specific-SFRs of cluster and field star-forming galaxies in bins of stellar mass (more precisely $K$-band absolute magnitude) are compared in Figure~\ref{ssfr_kmag}. For this analysis only those galaxies in the redshift range $0.15{<}z{<}0.25$ are considered, as at higher redshifts we become insensitive to the low specific-SFRs expected of galaxies in the process of being quenched (c.f. Fig.~\ref{ssfr_compare}). Each mass bin is fixed to contain the same number of galaxies (300), combining both cluster and field populations (irrespective of the relative contributions). The red shaded regions indicate the specific-SFR distribution of star-forming (24$\mic$-detected) cluster galaxies within $1.5\,r_{500}$ for each stellar mass bin. 
The graduations indicate the 1$\sigma$ (16--84\%) and interquartile (25--75\%) ranges, while the circle within the darkest shaded box indicates the median value and its uncertainty. 
The blue open boxes and symbols indicate the corresponding specific-SFR distributions of field galaxies in that same stellar mass bin. 

In agreement with the previous studies, the specific-SFRs of star-forming galaxies decline steadily with increasing stellar mass for both cluster and field samples. The steepness of this trend may be overestimated however, due to the observational bias that for a fixed SFR completeness limit, we can detect high mass star-forming galaxies to lower specific-SFRs than low-mass galaxies.

In each stellar mass bin, the distribution of specific-SFRs among cluster star-forming galaxies is ${\sim}3$0\% lower than that for the field galaxy counterparts. This reduction in specific-SFRs is seen in {\em all} stellar mass bins, and for {\em all} percentiles of the distribution shown (16, 25, 50, 75 and 84\%). This reduction in the specific-SFRs of star-forming galaxies in clusters must therefore be {\em systemic}. 
The values at the top of each mass bin indicate the significance (in units of $\sigma$) of the differences between the observed cluster and field specific-SFR distributions for that stellar mass bin, as determined by the non-parametric Mann-Whitney U-test.
These show that for four of the five bins, the specific-SFRs of star-forming cluster galaxies are lower than their field galaxy counterparts at the ${>}3{\sigma}$ level, the other being lower at the ${>}2{\sigma}$ level. 
Given the decrease in specific-SFR with stellar mass, some of this offset could in principle be produced by  mis-matches of the stellar mass distributions of cluster and field galaxies within each stellar mass bin. This possibility can be excluded, as the mean stellar masses  ($\langle\mathcal{M}\rangle$) of the cluster and field sub-samples within each bin are consistent with one another to within 0.01\,dex for all five of the stellar mass bins.

We also checked the possibility that the lower specific-SFRs of the cluster galaxies are due to them being at slightly lower redshifts than their field counterparts. While we find that the cluster galaxies in Figure~\ref{ssfr_kmag} have mean redshifts 0.003--0.006 lower than the field galaxies from the same stellar mass bin, based on the rate of evolution reported in Fig.~\ref{ssfr_evol} and Table~\ref{table:ssfr}, this should produce an offset of just 0.0096--0.0195\,dex (2--5\%) in their specific-SFRs, much lower than the ${\sim}3$0\% reduction observed. We will also show in Section~\ref{sec:slowquench} that the specific-SFRs of star-forming cluster galaxies are ${\sim}3$0\% lower than matched star-forming field galaxies of the same stellar mass {\em and} redshift.
This result shows conclusively that the SFRs of star-forming galaxies in clusters are systematically reduced with respect to the field at fixed stellar mass. Moreover, this reduction in the SFRs is occurring at all stellar masses (at least for $\mathcal{M}{\ga}10^{10}{\rm M}_{\odot}$). 
As discussed above, this systemic reduction of SFRs among star-forming cluster galaxies is indicative of them being slowly quenched on ${\ga}1$\,Gyr time-scales upon being accreted into the cluster. 

\begin{figure}
\centerline{\includegraphics[width=84mm]{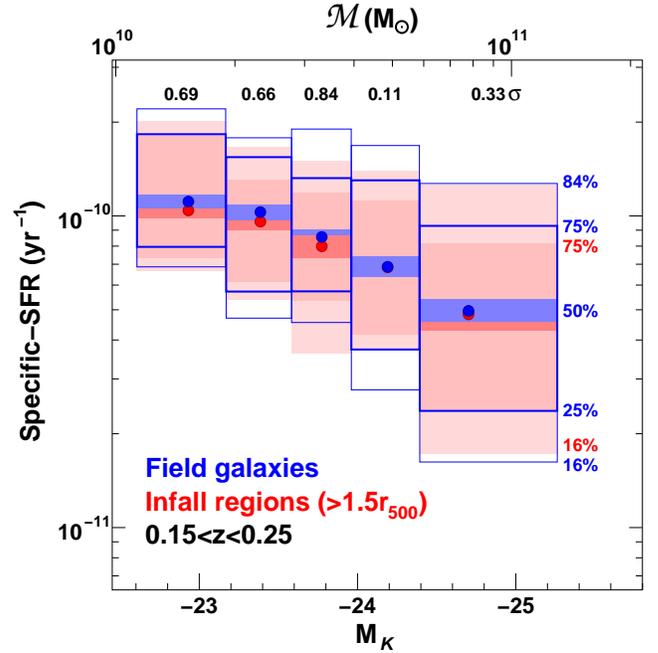}}
\caption{Comparison of the main sequence of star-forming galaxies in the infall regions of clusters ($r_{proj}{>}1.5\,r_{500}$) and the field at $0.15{<}z{<}0.25$. The different colors, symbols and shaded regions are as in Fig.~\ref{ssfr_kmag}. Each bin in stellar mass now contains 224 galaxies combining both cluster and field galaxies. }
\label{ssfr_kmag2}
\end{figure}

To confirm that it is the arrival into the cluster that is inducing the quenching of star formation in these galaxies, rather than environmental processes active at larger cluster-centric radii, such as pre-processing, we repeated the above analysis, but this time comparing galaxies in the cluster infall regions outside $1.5\,r_{500}$ with those from the field, the results of which are shown in Figure~\ref{ssfr_kmag2}. 
The specific-SFRs of star-forming galaxies in the infall regions of clusters (${>}1.5\,r_{500}$) are indistinguishable from those in the general field for all five stellar mass bins. This suggests that the slow quenching of star-forming galaxies revealed by Fig.~\ref{ssfr_kmag} must be confined to the cluster itself (${\la}1.5\,r_{500}$) and not extend into the infall regions. This does not rule out quenching of any form at these large radii, but if quenching is occurring outside $1.5\,r_{500}$, it must be too rapid to leave any imprint on the specific-SFR distribution of the infalling star-forming galaxies. We repeated this analysis, adjusting the boundary between cluster and infall regions from the value of $1.5\,r_{500}{\simeq}r_{200}$ presented here in Figures~\ref{ssfr_kmag} and~\ref{ssfr_kmag2}, but this proved to be the cluster-centric radius at which the two behaviors were best separated. 

\section{The slow quenching of star formation in cluster galaxies}
\label{sec:slowquench}

If star forming galaxies at the present day are affected by environmental mechanisms when they move from the isolated field to become bound within groups or clusters, we should see a signature of this transformation which depends on the time-scale over which this transformation occurs. As discussed in Section~\ref{sec:quenching}, comparing the SFRs of star-forming galaxies in cluster and field regions allows constraints on the quenching time-scale to be made. Increasing the time-scale over which star formation is quenched upon accretion into a massive system will increase the numbers of star-forming galaxies in clusters being observed during this quenching process, thus skewing the overall SFR and specific-SFR distributions of star-forming cluster galaxies to lower values than their counterparts in the field. 

 The results presented in Figs~\ref{ssfr_kmag} and~\ref{ssfr_kmag2} suggest a simple scenario in which infalling star-forming field galaxies are continually accreted into a massive cluster, forming stars normally until they pass within $1.5\,r_{500}{\approx}1.0\,r_{200}$, whereupon they are slowly quenched by some cluster-related mechanism, systematically reducing their SFRs. No obvious stellar mass dependence for the quenching is apparent in Fig.~\ref{ssfr_kmag}, suggesting that all star-forming galaxies with $\mathcal{M}{\ga}10^{10}M_{\odot}$ are affected in the same manner. 

\begin{figure*}
\centerline{\includegraphics[width=170mm]{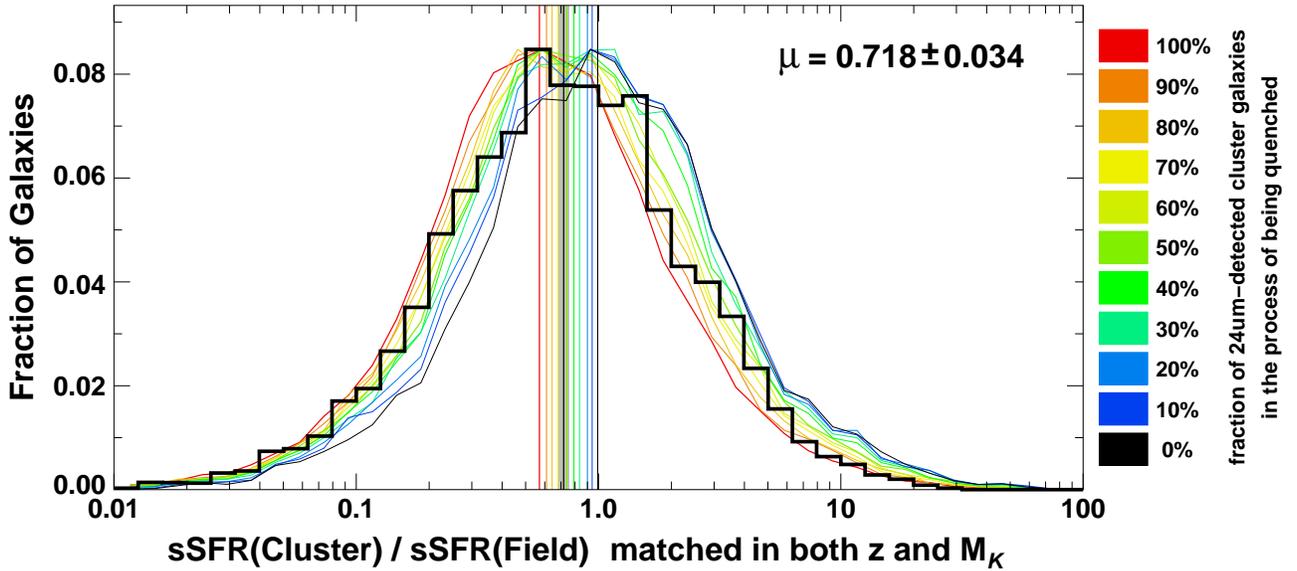}}
\caption{Comparison of the specific-SFRs of cluster star-forming galaxies and matched field star-forming galaxies of the same stellar mass and redshift. The thick black histogram shows the population distribution of sSFR(cluster)/sSFR(field) for random pairs of cluster and field star forming galaxies matched in stellar mass and redshift. The vertical black line and shaded region shows the median value of sSFR(cluster)/sSFR(field) and its 1$\sigma$ confidence limit. The thin colored histograms show the corresponding distributions of sSFR(mixed population; $f_{Q}$)/sSFR(field), where mixed populations consist of a fraction $f_{Q}$ of quenching galaxies added to a fraction $(1-f_{Q})$ of the original star-forming field galaxy population. Eleven curves indicate the distributions for values of $f_{Q}$ from 0\% (black) to 100\% (red), color coded as indicated to the right of the plot. The median value for each mixed population distribution is indicated by the correspondingly colored vertical line.}
\label{quenching}
\end{figure*}

\subsection{How many cluster galaxies are in the process of being quenched?}
\label{sec:modelquench}

We do not expect to be witnessing all of these star-forming (i.e. 24$\mu$m-detected) cluster galaxies at exactly the same point in their transformation from normal star-forming field galaxies into passively-evolving cluster members. 
We are rather seeing them at various stages of the quenching process, prior to their SFRs falling below the threshold required for them to be detected in our 24$\mic$ images, while others have not yet encountered the cluster environment and had their SFRs affected. 

To gain insight into this quenching mechanism, we first need to estimate the fraction of star-forming cluster galaxies which are observed in the process of being quenched, $f_{Q}$, as opposed to forming stars normally. To do this, we compare the SFRs of cluster and field star-forming galaxies at fixed stellar mass and redshift, and attempt to reproduce the observed differences with synthetic mixed population models containing field and quenching populations.

For each star-forming cluster galaxy ($r_{proj}{<}1.5\,r_{500}$; $M_{K}{<}{-}23.1$, $0.15{<}z{<}0.25$), we match it at random with a star-forming field galaxy with the same stellar mass (within 0.02\,dex) and redshift (${\delta}{z}{<}0.02$), and compare their specific-SFRs (sSFR). This is repeated for all cluster galaxies, and multiple times so that all possible matches are included, building up the distribution of sSFR(Cluster galaxy)/sSFR(Field galaxy). 
As each of the five stellar mass bins in Fig~\ref{ssfr_kmag} show similar offsets between the cluster and field specific-SFR distributions, we can combine all five bins to estimate a single value of $f_{Q}$ for all $M_{K}{<}{-}23.1$ star-forming cluster galaxies. 

The resulting distribution of sSFR(Cluster)/ sSFR(Field) is shown as the thick black histogram in Figure~\ref{quenching}. An approximately log-normal distribution is obtained, centered close to unity, but slightly offset to the left (lower specific-SFRs in cluster galaxies) and with a flattened top. Few matched pairs show ratios below 0.1 or greater than 10, due to a combination of the limited width of the specific-SFR distribution for the main sequence of star-forming galaxies at fixed stellar mass, and the relative rarity of galaxies at $0.15{<}z{<}0.30$ with 24$\mic$ fluxes ${\ga}10{\times}$ above the completeness limit. The median value of sSFR(Cluster)/sSFR(Field) for a single matched sample of cluster and field star-forming galaxies is found to be  ${\mu}{=}0.718{\pm}0.034$, and is marked by the vertical black line and shaded region indicating the 1$\sigma$ confidence range. In other words, the specific-SFRs of cluster star-forming galaxies are 28\% lower than their field counterparts at fixed stellar mass and redshift. This reduction is significant at the 8.7$\sigma$ level.

For simplicity, and ease of comparison to other studies \citep[e.g.][]{wetzel}, we consider a quenching model in which the SFRs of star-forming galaxies decline exponentially upon arrival in the cluster, with e-folding quenching time-scale $t_{Q}$, such that ${\rm SFR}(t){=}{\rm SFR}(0)\exp(-t/t_{Q})$. To create artificial quenching populations, we take multiple copies of the observed star-forming field galaxy population, and for each subsequent copy reduce the SFR and 24$\mic$ flux to match this slow exponential decline in star formation, until the last copy represents the cluster population after 10\,$t_{Q}$ (the actual length of this time-scale is unimportant here), by which point none of them should remain detectable at 24$\mic$. This synthetic quenching population thus contains galaxies which started out as normal star forming galaxies, but have since been undergoing quenching for a time $t$, where $t$ is uniformly distributed between 0 and 10\,$t_{Q}$. The model thus assumes that the triggering of this slow quenching process has been affecting galaxies at a uniform rate for the last ten $t_{Q}$. 
We then remove all those {\em quenching} field galaxies whose 24$\mic$ emission now lies below our 24$\mic$ completeness limit. This naturally removes more of the galaxies which had been undergoing quenching for longer. 
What remains approximates that component of the star-forming galaxy population observed at all stages of the slow quenching process, but which still would be detectable in our 24$\mic$ images. These 24$\mu$m-detectable quenching galaxies have on average been undergoing quenching for a median of 0.95\,$t_{Q}$. They constitute just 15\% of the model quenching galaxies created. In other words, the number of 24$\mu$m-detected quenching model galaxies is equal to the total number of star-forming galaxies (24$\mu$m-detected at $t{=}0$) fed into the model (accreted into the cluster) within the last 1.49\,$t_{Q}$.
These values (0.95\,$t_{Q}$, 1.49\,$t_{Q}$) depend solely on the 24$\mu$m distribution of the observed star-forming field galaxies with respect to our 24$\mu$m completeness limit. Galaxies which start out with 24$\mu$m fluxes only marginally above the completeness limit will rapidly become undetectable after beginning the quenching process, while others that begin with much higher 24$\mu$m fluxes will continue to remain detectable for several quenching time-scales.  
 
We create mixed populations combining a fraction $f_{Q}$ of galaxies from the synthetic quenching population and $1{-}f_{Q}$ from the original star-forming field galaxy population.
These mixed populations are matched with galaxies from the original star-forming field population with the same stellar mass and redshift (excluding matches involving copies of the same field galaxy) in the same way as before, to produce the distributions of sSFR(mixed population)/sSFR(Field) for values of $f_{Q}$ ranging from 0 to 1 shown by the thin colored curves in Figure~\ref{quenching}.  

As expected, the histogram for $f_{Q}{=}0$ ({\em thin black curve}) is centered at 1.0, as this is matching the original field star-forming galaxy sample with itself. As $f_{Q}$ increases, the entire distribution of sSFR(mixed population)/sSFR(Field) shifts to lower values. However, even when $f_{Q}{=}1$ ({\em red curve}), i.e. all of the star-forming galaxies are observed in the process of being quenched, the median reduction in the specific-SFR is just ${\sim}4$5\%. This is simply due to the limited range of 24$\mu$m fluxes in the original field galaxy sample. 

Comparing the median value of sSFR(Cluster)/ sSFR(Field) of $0.718{\pm}0.034$ with that from our mixed quenching populations, we estimate that ${\sim}60{\pm}10$\% of the star-forming cluster galaxies within a {\em projected} cluster-centric radius of $1.5\,r_{500}$ are in the process of being quenched when observed. The overall distribution of sSFR(Cluster)/sSFR(Field) shown by the thick black histogram also appears consistent with an evenly mixed population of quenching and non-quenching galaxies, following approximately the green curves and consistently between the extreme red and black ones.  
 
\subsection{Estimating the quenching time-scale $t_{Q}$}
\label{timescale}

In the previous section, we estimated that ${\sim}60$\% of the star-forming cluster galaxies within $1.5\,r_{500}$ are in the process of being quenched when observed. 
While this informs us of the likely size of the quenching population (we estimate that ${\sim}500$ of our 24$\mu$m-detected spectroscopic cluster members are observed in this process of slow quenching, or ${\sim}11$\% of all $M_{K}{<}-23.1$ cluster members within $1.5\,r_{500}$), this does not by itself constrain the e-folding time-scale $t_{Q}$ over which this quenching is taking place on average. To do this we need to place these clusters in the cosmological context of continually growing by accreting galaxies and groups from the surrounding large-scale structure. 
We can achieve this by using cosmological simulations to firstly estimate the rate at which clusters are accreting galaxies from their surroundings at these redshifts ($z{\sim}0.2$), and secondly estimate the likely fractions of galaxies which would be spectroscopically identified as cluster members but are in fact physically outside $r_{200}$.

In observations of distant clusters with velocity dispersions of the order 1\,000\,km\,s$^{-1}$ it is not possible to identify whether any particular galaxy is physically located within 
$1.5\,r_{500}$ of the cluster center, or rather much further out and hence has yet to encounter the cluster environment. However, by comparison to cosmological simulations containing dozens of massive clusters similar to those from LoCuSS, and populated by galaxies based on semi-analytic models, it is possible to build simulated ``observed'' caustic diagrams of massive clusters for which the accretion histories of the member galaxies are known.  
In \citet{haines12} we extracted regions centered on the 30 most massive clusters from the Millennium Simulation \citep{springel_mill}, a cosmological dark matter simulation covering a (500\,$h^{-1}$\,Mpc)$^{3}$ volume. For each galaxy within this region, there exists a full database of its properties including positions, peculiar velocities, absolute magnitudes etc. based upon the semi-analytic models ({\sc galform}) of \citet{bower}, at 63 snapshots throughout the lifetime of the Universe. This allows the orbit of each galaxy around the cluster to be followed from its formation right up to the present day. We could then build stacked caustic diagrams for clusters observed at $z{=}0.21$ in which each galaxy is coded according to the epoch at which it was accreted into the cluster \citep[see Fig.~3 of][]{haines12}. We use the same analysis here, except now we extend the sample to the 75 most massive clusters to better sample some of the lower mass systems in LoCuSS, and fix the epoch of accretion into the cluster as the redshift at which the galaxy passes within $r_{200}$ for the first time. 

Stacking over these 75 clusters, we find that 15.3\% of $M_{K}{<}{-}23.1$ galaxies with $r_{proj}{<}r_{200}$ and line-of-sight velocities placing them within the caustics of that cluster, and hence would be identified as spectroscopic cluster members, are actually ``interlopers'', galaxies which are physically located outside $r_{200}$ at the time of observation ($z{=}0.21$) and which have not been inside this radius at any prior epoch. If we assume that the $40{\pm}10$\% of star forming cluster galaxies forming stars normally all form part of this 15.3\% ``interloper'' population, then we associate the remaining $60{\pm}10$\% which are currently undergoing quenching with the 22.9$_{-7.6}^{+12.7}$\% of cluster galaxies which have been most recently accreted. This makes no assumption on the actual $f_{SF}$ of these interloping field galaxies, except that the $f_{SF}$ among those galaxies at the moment of accretion (passing within $r_{200}$) is the same as that of the whole interloper population. As these are mostly galaxies from the immediate infall regions, this should be a reasonable approximation. Interestingly, if we assume the fraction of star-forming galaxies among these interlopers to be the same as that of field galaxies at these redshifts ($f_{SF,field}{\sim}2$5\%), we can estimate the contribution to the $f_{SF}$ of our clusters just from these interlopers ($0.153{\times}f_{SF,field}$), as indicated by the gray shaded region in Fig.~\ref{boevol}. This estimate suggests that ${\sim}5$0--60\% of the cluster galaxies classed as star-forming are in fact at $r{>}r_{200}$ and have yet to encounter the cluster environment, consistent with the model described above. 

The 22.9$_{-7.6}^{+12.7}$\% of ``spectroscopic'' members which are the most recent arrivals into the 75 clusters were all accreted within the $1.74_{-0.67}^{+1.21}$\,Gyr leading up to $z{=}0.21$, i.e. after $z{\sim}0.40$. As discussed above, this equates to 1.49\,$t_{Q}$, implying a quenching time-scale of $t_{Q}{=}1.17_{-0.45}^{+0.81}$\,Gyr, i.e. in the range 0.72--1.98\,Gyr. Here the only source of uncertainty we include comes from the confidence limits in $f_{Q}$. We note that the key source of uncertainty not included here comes from the choice of model itself, and our assumption for the cluster-centric radius at which quenching is switched on. For example, reducing the radius at which quenching is initiated from $r_{200}$ to $r_{500}$, the resultant quenching time-scale doubles to 2.5\,Gyr. 
These time-scales are consistent with \citet{moran07} who found significant populations of passive spirals as well as E/S0s in two $z{\sim}0.4$ clusters whose UV colors and spectral indices (d4000) indicate that their SFRs have been declining on ${\sim}1$\,Gyr time-scales. 

\section{Discussion}
\label{sec:discussion}

\subsection{The origin of the Butcher--Oemler effect}
\label{sec:discussbo}

The Butcher--Oemler effect has remained one of the most enduring topics of discussion within the field of galaxy evolution over the last 30 years. 
In Fig~\ref{bo}, we charted the rapid rise in the fraction of cluster members within $1.5\,r_{500}$ with active star formation (${\rm SFRs}{>}3\,{\rm M}_{\odot}$\,yr$^{-1}$) from $f_{SF}{=}0.023{\pm}0.013$ at $z{\sim}0$ to $f_{SF}{=}0.185{\pm}0.025$ at $z{\sim}0.40$ --- the mid-IR Butcher--Oemler effect. It is notable that the evolutionary trend we obtain is essentially the same as that from the original analysis of BO84. This is perhaps surprising given the numerous selection biases \citep[see e.g.][]{andreon99,depropris03,holden,haines09a} that are known to affect the BO84 analysis, such as selecting galaxies on optical luminosity ($M_{V}$) rather than stellar mass, and identifying star-forming galaxies by their blue optical colors and hence missing the numerous dusty star-forming galaxies found within the cluster red sequence \citep{wolf05,haines08}. The key gains however of our new analysis is that, being entirely based upon spectroscopically-confirmed cluster members for which star-forming galaxies are selected according to robust 24$\mu$m-derived SFR measurements (rather than simple red/blue color splits), we can now dissect the Butcher--Oemler effect into its various evolutionary constituents and quantify their relative contributions. 

In Section~\ref{sec:boevol} we showed that a significant component of the Butcher--Oemler effect is simply due to the cosmic decline in star-formation among all field galaxies, which are subsequently accreted onto the clusters, rather than any cluster-specific processes. This is confirmed by our finding that, while the infrared LFs of cluster galaxies have evolved rapidly since $z{\sim}0.8$ (Fig.~\ref{lf_evol}), at all redshifts they are indistinguishable from the LFs of coeval field galaxies. 
However, even after accounting for the cosmic evolution among star-forming field galaxies of the form $L_{TIR}(z){\propto}(1+z)^{3.4}$ \citep{rujopakarn}, a significant Butcher--Oemler effect of the form $f_{SF}{\propto}(1+z)^{2.78}$ (Fig~\ref{boevol}) was found to persist, implying that additional suppression of star formation due to cluster-specific processes is required. This was confirmed in Fig.~\ref{popesso} which showed that the total star formation activity per unit halo mass has declined much more rapidly since $z{\sim}0.3$ in clusters than the field. 

As discussed in Section~\ref{sec:ssfr}, the cosmic decline in star formation since $z{\sim}1$ is best described as a systematic reduction in the specific-SFRs of star forming galaxies at all masses \citep{noeske,elbaz}. 
Figure~\ref{ssfr_evol} charted the evolution in the specific-SFR distributions of cluster and field galaxies over $0.0{<}z{<}0.30$, revealing that the decline in the specific-SFRs among cluster galaxies is more rapid than that seen for field galaxies in the same {\em Spitzer}/MIPS images, as well as that implied by the $(1+z)^{3.4{\pm}0.2}$ evolution of star-forming field galaxies observed by \citet{rujopakarn}. The factor $2.78{\pm}0.24$ decline in the specific-SFRs of cluster galaxies observed going from our high redshift bin ($0.25{<}z{<}0.30$) to the local cluster sample ($0.023{<}z{\le}0.060$) is instead best fit by an evolutionary model of the form $(1+z)^{5.02{\pm}0.44}$.

Reperforming the Butcher--Oemler analysis of Section~\ref{sec:bo}, but this time having an evolving SFR threshold of the form $(1+z)^{5.02}$ to account for our observed luminosity evolution, we find that even with this extremely rapid luminosity evolution, a small residual Butcher-Oemler effect remains, with $f_{SF}{\propto}(1+z)^{2.18{\pm}1.16}$. Excluding the two $z{\sim}0.4$ clusters from the fit flattens the residual trend to $f_{SF}{\propto}(1+z)^{0.66{\pm}0.94}$. 
This implies a factor 1.19--1.77 increase in the fraction of star forming galaxies in clusters from the present day to $z{=}0.3$ over and above the observed luminosity evolution, or a factor 1.25--2.08 increase by $z{=}0.40$. 
This ${\sim}2{\times}$ density evolution is consistent with the predicted evolution in the rate at which galaxies are accreted into massive clusters over this time period. From cosmological simulations of massive clusters, the {\em fraction} of member galaxies accreted within the previous Gyr is found to approximately double from $z{\sim}0$ to $z{\sim}0.4$ \citep{berrier}. Alternatively, this density evolution could reflect a reduction in the $f_{SF}$ among those field galaxies accreted onto the clusters since $z{\sim}0.4$, perhaps due to pre-processing in groups (McG09), or being quenched by secular processes \citep[``mass quenching'';][]{peng12} in the intervening period. 

We have identified three main components which combine to produce the overall decline in star formation activity among cluster galaxies of the form $f_{SF}{\propto}(1+z)^{6.26}$ since $z{\sim}0.3$: (i) the cosmic decline in star formation among field galaxies which are subsequently accreted into the clusters (55\% contribution); (ii) an {\em accelerated} decline in the specific-SFRs of cluster galaxies (30\%); and (iii) a reduction in the numbers of star-forming cluster galaxies (15\%). 
As this accelerated decline in specific-SFRs among cluster galaxies was revealed in Fig.~\ref{ssfr_compare} by the systematically lower specific-SFRs among star-forming cluster galaxies than their counterparts in the field at $0.15{<}z{<}0.25$, and this same observation is evidence that star-formation is being slowly quenched in cluster galaxies, we believe that this accelerated decline in star formation activity is due to this slow quenching process. 

We note that slow quenching does not necessarily produce an accelerated decline in $f_{SF}$ (c.f. McG11). If it affects the same fraction of star-forming cluster galaxies at each redshift, and by the same amount, then the $f_{SF}$ would be reduced by the same factor at all redshifts. An accelerated decline in the $f_{SF}$ could suggest that the mechanism behind the slow quenching process is becoming more effective with time. This could occur by shortening the quenching time-scale, $t_{Q}$, (which we have effectively only measured at $z{\sim}0.2$) resulting in fewer star-forming galaxies remaining in present day clusters. Alternatively, the range over which the quenching mechanism is effective could increase with time, say from $r_{500}$ at $z{\sim}0.4$, to $1.5\,r_{500}$ at $z{\sim}0.2$ and $2\,r_{500}$ at $z{=}0.0$. This would have the effect of increasing the fraction of star-forming cluster galaxies within $1.5\,r_{500}$ observed while undergoing quenching, $f_{Q}$, at later epochs, as well as reducing their overall numbers.  

Such an increased effectiveness of environmental quenching processes with time may explain the trends revealed in Figure~\ref{popesso}, whereby the star formation activity of cluster galaxies has been declining much more rapidly since $z{\sim}0.3$ than over $z{\sim}0$.3--1.0, when it fell approximately in step with the steady cosmic decline in star formation among field galaxies. 

The increasing impact of environmental processes with time can be plausibly explained within ram-pressure stripping or starvation scenarios. While the ram pressures felt by galaxies infalling into massive clusters are not predicted to evolve significantly between $z{\sim}1$ and the present day \citep[e.g.][]{bruggen,tecce}, assuming the original \citet{gunn} condition of $P_{ram}{>}2{\pi}G \Sigma_{star} \Sigma_{gas}$, galaxies should become increasingly susceptible to ram-pressure stripping as their gas surface density $\Sigma_{gas}$ falls. Given the Schmidt-Kennicutt SFR scaling relation of $\Sigma_{SFR}{\propto}\Sigma_{gas}^{N}$ where $N{\sim}1$.0--1.5 \citep{kennicutt,bigiel}, tightly linking the density of star formation to gas density, the $\Sigma_{gas}$ of normal star-forming spirals should evolve roughly in parallel to the observed ${\sim}10{\times}$ decline in star formation since $z{\sim}1$ \citep[assuming the sizes of disks has not evolved signficantly over this period;][]{barden}. 
This rapid decline in $\Sigma_{gas}$ since $z{\sim}1$ has been confirmed by recent CO observations of normal star-forming spirals at $z{\sim}1$.0--1.5 \citep{daddi,genzel}. Spirals should thus become more vulnerable to ram-pressure stripping at low redshifts.

\subsection{The slow quenching of star formation in cluster galaxies}

A primary driver for many studies focused on understanding the origins of the SF--density relation is to establish both the range of environments where star-formation in cluster or group galaxies is quenched (e.g. minimum halo mass, M$_{min,Q}$, maximum cluster-centric radius in units of $r_{200}$) and the average time-scales over which star formation is quenched upon arrival in these environments \citep[McG09, McG11;][hereafter DeL12]{balogh10,delucia}. These two key constraints allow us in principle to distinguish among the various physical mechanisms proposed to transform galaxies in dense environments \citep[e.g.][]{treu}. 

In the case of rapid quenching ($t_{Q}{\la}100$\,Myr), the resulting unmistakable spectral imprint of deep Balmer absorption lines combined with weak or absent emission lines, should allow recently quenched galaxies to be identified up to ${\sim}1$\,Gyr after the truncation event \citep{leonardi}. Spectroscopic surveys of cluster galaxies have however found such ``post-starburst'' galaxies to be essentially absent in local clusters, at least among ${\ga}L^{*}$ galaxies \citep{poggianti04,yan}, largely ruling out such short quenching time-scales. 

The focus has thus switched to finding evidence of gentle environmental processes which quench star formation in cluster galaxies on much longer time-scales \citep[${\ga}1$\,Gyr, e.g.][]{moran07,wetzel}. 
The most characteristic signature of ongoing slow quenching occurring within cluster or group galaxies is an increased frequency of galaxies showing either intermediate UV--optical colors placing them between the cluster red sequence and blue cloud populations, or equivalently reduced specific-SFRs in comparison to star-forming field galaxies of the same redshift and stellar mass.  

The clearest evidence for this transition population was seen in Figure~\ref{ssfr_kmag}, which compared the specific-SFR distributions of cluster and field galaxies in fixed bins of stellar mass. In each stellar mass bin, the specific-SFRs of star forming cluster galaxies within $1.5\,r_{500}$ were found to be systematically ${\sim}3$0\% lower than their counterparts in the field. Considering all $M_{K}{<}{-}23.1$ galaxies, this systematic reduction in the SFRs of cluster galaxies (at fixed stellar mass and redshift)
was found to be significant at the 8.7$\sigma$ level ({\S}~\ref{sec:modelquench}). 
This figure provides unambiguous evidence that star formation in most (and possibly all) massive ($\mathcal{M}{\ga}10^{10}{\rm M}_{\odot}$) star-forming galaxies is being {\em slowly} quenched once they are accreted by a massive cluster, by revealing the slow, continual and inexorable suppression of their star formation activity.

Figure~\ref{ssfr_kmag2} suggests that this slow quenching process is confined to star-forming galaxies within $1.5\,r_{500}$, and so we developed a simple model in which star-forming galaxies are continually accreted into a massive cluster, forming stars normally until they pass within $1.5\,r_{500}{\approx}r_{200}$, whereupon they are slowly quenched, their SFRs declining exponentially with an e-folding time-scale $t_{Q}$. Figure~\ref{quenching} showed that this simple model was able to reproduce both the level and form of the systematic reduction in the specific-SFRs displayed by star-forming cluster galaxies. The best-fitting model suggested that of the star-forming cluster galaxies located within a projected cluster-centric distance of $1.5\,r_{500}$ and detected at 24$\mu$m, some 40\% are forming stars normally, and hence in this model are identified as being physically located beyond $1.5\,r_{500}$, while the remaining 60\% are observed in the process of being slowly quenched, having recently been accreted by the cluster by passing within $1.5\,r_{500}$ for the first time. By comparing these two populations to realistic ``observations'' of 75 massive clusters drawn from the Millennium simulation, for which detailed galaxy accretion histories were derived, we were able to estimate an average quenching time-scale of $t_{Q}{=}1.17_{-0.45}^{+0.81}$\,Gyr. 

A quenching time-scale of 1--2\,Gyr ultimately appears most likely for two reasons: (i) if it was much shorter, there would not be sufficient {\em quenching} galaxies to skew the specific-SFR distribution of star forming galaxies in clusters as much as we observe; (ii) if $t_{Q}$ was much longer, then given that half of all cluster galaxies are accreted in the last 4--5\,Gyr the quenching process now becomes too slow to shut down star formation in sufficient numbers of these late arriving field galaxies to leave local clusters with as few star forming galaxies as observed (Fig.~\ref{bo}). Furthermore, if the quenching time-scale is much longer than the cluster-crossing time-scale (${\sim}2$\,Gyr), then the resulting radial population gradients would also be much shallower than observed \citep[Fig.~7]{haines09a}. We will return to the question of estimating the quenching time-scale in the next paper, using a dynamical analysis and the radial population gradients to provide two further independent estimates of $t_{Q}$. 

The slow quenching of star formation in galaxies upon arrival in a cluster on 1--2\,Gyr time-scales is exactly what is predicted for starvation models, in which galaxies are rapidly stripped of their diffuse gaseous halos via their passage through the ICM, preventing further gas accretion onto the galaxies from the surrounding inter-galactic medium \citep{larson,bekki,mccarthy}. The galaxy then slowly uses up its existing molecular gas reservoir through star formation over a period of ${\sim}2.3$\,Gyr, based on the constant molecular gas consumption time-scales observed for the disks of nearby spiral galaxies spanning a wide range of properties \citep{bigiel}.

Ram-pressure stripping is often viewed as a much more rapid, violent process than starvation, and one that is also limited to cluster cores. 
Hydrodynamical simulations confirm that the gas contents of massive spiral galaxies are stripped on ${\la}1$00\,Myr time-scales via a combination of ram-pressure and viscous stripping when subjected to winds comparable to a high-velocity passage through the dense ICM of the cluster core \citep{quilis}. Such short time-scales though appear inconsistent with the frequent finding in nearby clusters of partially ram-pressure stripped spirals with truncated gas and H$\alpha$ disks, and outer regions showing recently quenched stellar populations \citep[e.g.][]{cayatte,vogt,koopmann,crowl}. Moreover, evidence for ongoing ram-pressure stripping is seen for spiral galaxies at distances of ${\sim}1$\,Mpc or ${\sim}r_{500}$ from the cluster center \citep[e.g.][]{chung,merluzzi}, indicating that spirals are being affected by ram-pressure or viscous stripping well outside the cluster cores. Hydrodynamical simulations confirm that even in the low-density regions of the cluster outskirts (${\sim}r_{500}-r_{200}$) the moderate ram pressures ($P_{ram}{\sim}1$00--1000\,cm$^{-3}$\,km$^{2}$\,s$^{-2}$) acting on infalling massive spirals are sufficient to truncate their gas disks and strip half of their cold gas contents over 500--1000\,Myr periods \citep{roediger05}, resulting in commensurate reductions in their SFRs. \citet{bahe} find that the ram pressures felt by infalling spirals are sufficient to begin stripping their cold gas and affect their star formation at $r_{200}$, but their hot gas atmospheres can be stripped out to ${\sim}5\,r_{200}$. 

Gas-rich spiral galaxies falling into clusters for the first time do not suddenly find themselves in the cluster centers encountering the dense ICM and the high ram pressures capable of stripping all of their gas on ${\la}1$00\,Myr time-scales.  
Instead they encounter gradually increasing ICM densities and ram-pressures \citep[see Fig.~3 of][]{bruggen} which incrementally strip gas from their disks from the outside-in as they travel inwards from the cluster outskirts to the pericenter of their orbit in or near the cluster core. The effective time-scale for quenching via ram-pressure stripping then becomes the ${\sim}1$\,Gyr time it takes for infalling spirals to get from the cluster outskirts where they first encounter the ICM to the cluster core where the final stages of gas stripping are completed. This is shown in the hydrodynamical simulations of \citet{roediger} which follow the effects of ram-pressure stripping on spirals following realistic orbits in a rich cluster (see their Figure 7).

\subsection{Comparison to previous studies}

The results presented here that support the predominant role of slow quenching of star-formation in cluster galaxies appear inconsistent with \citet{balogh04}, who first performed this form of analysis and found the EW(H$\alpha$) distribution of local star-forming galaxies (from the SDSS and 2dFGRS) to be independent of local density. They concluded that the environmental quenching of star formation must be a rapid process, preferentially taking place at high redshifts. 
Subsequent similar analyses of local bright ($M_{r}{<}M_{r}^{*}{+}1$) star forming galaxies from the SDSS \citep{tanaka,haines07} and GAMA surveys \citep{wijesinghe} again found no dependence of the EW(H$\alpha$) or SFR distributions (at fixed stellar mass) on local density. We believe that this apparent contradiction is due to two key differences in the methodology and type of survey involved. 

Firstly, while we are focused solely on the immediate environs of massive clusters, the previous studies were volume-limited surveys covering the full range of environments with few (if any) rich clusters included. Instead, the galaxies in their high-density regions are predominately associated to galaxy groups rather than clusters, simply reflecting the observation that more than half of galaxies within the local Universe are found in groups, while just a few percent are associated to massive clusters (McG09). Hence their SFR--density relations are defined primarily by environmental processes active within galaxy groups of mass ${\sim}10^{13}{\rm M}_{\odot}$.
Secondly, \citet{balogh04}, \citet{tanaka} and \citet{wijesinghe} all use $\Sigma_{5}$ as a measure of local density, based on the distance to the fifth nearest $M_{r}{<}{-}20$ galaxy within 1\,000\,km\,s$^{-1}$, which as \citet{haines07} show is rather insensitive to the location of a galaxy within a galaxy group. While galaxies with high $\Sigma_{5}$ are almost all members of bound systems, the converse is not true, with group members found across the full range of $\Sigma_{5}$. Any systematic suppression in the SFRs of galaxies within groups or clusters will thus be smeared out when viewed in terms of the SFR--$\Sigma_{5}$ relation. In contrast, by defining our environment in terms of $r/r_{500}$, we are able to directly link the properties of the galaxies to their location within the DM halo of a massive cluster (after accounting for projection effects) or outside it, and measure the related impact of being accreted into the cluster halo on their star formation activity, something which cannot be achieved using $\Sigma_5$ alone. 
 
It is this combination of the type of survey used (cluster-specific versus representative volume of the Universe) and the metric used to define local environment ($r/r_{500}$, versus local density, $\Sigma_{5}$) that likely lie behind the apparently inconsistent results presented here and in the previous studies described above. Our results are also in agreement with \citet{vonderlinden} who analysed the star formation activity of galaxies as a function of cluster-centric radius for 521 clusters at $z{<}0.1$ using SDSS spectroscopic data. Like us, they found that within $r_{200}$ the SFRs and specific-SFRs of star-forming cluster galaxies systematically decline towards the cluster core from the values seen in the field, while \citet{boselli} found similar trends for late-type galaxies in the vicinity of both Coma and Virgo clusters. 

By directly comparing the specific-SFRs of star-forming galaxies in clusters and the field at fixed redshift and stellar mass (Figs~\ref{ssfr_kmag} and~\ref{quenching}), we have been able to effectively isolate and quantify the impact of the cluster environment on star formation, having minimized any contribution from evolutionary or mass-dependent effects. While our estimates for $t_{Q}$ are model dependent, the actual evidence for slow quenching in cluster galaxies is a clean, model-free result. 

The 0.7--2.0\,Gyr range for $t_{Q}$ we obtain here is consistent with recent estimates of ${\sim}3$\,Gyr for the {\em total} time required to transform an infalling star-forming galaxy into a passively-evolving (${\rm sSFR}{<}10^{-11}{\rm yr}^{-1}$) cluster member \citep[McG09, McG11;][DeL12]{balogh10}. We note that none of the latter estimates are based on direct observation of a {\em quenching} galaxy population within clusters, but rather by comparing the evolution or scatter in the fraction of passive/red galaxies ($f_{pass}$) within groups and clusters, with the predicted galaxy accretion histories of clusters obtained from cosmological simulations. 
In each case, these estimates assume that galaxies have become passive {\em solely} through environmental processes related to their accretion into a massive halo. 
McG09 and McG11 argue that the rapid increase in $f_{pass}$ among group and cluster galaxies since $z{\sim}0.4$ is consistent with slow quenching on ${\sim}3$\,Gyr time-scales after being accreted into a ${\sim}10^{13}{\rm M}_{\odot}$ halo. 
However, as we discuss in {\S}~\ref{sec:discussbo}, a significant fraction of the decrease in $f_{SF}$ in clusters since $z{\sim}0.4$ is due to the cosmic decline in star formation in all galaxies, while many cluster galaxies are also likely to have become passive due to internal mass-related mechanisms \citep{bundy}, such as AGN feedback \citep[e.g.][]{bower}. These secular processes are also likely to contribute significantly to the low scatter in $f_{pass}$ seen among groups and clusters by shutting down star formation in most massive galaxies irrespective of environment, and could plausibly explain the estimated values of $t_{Q}{\sim}3$\,Gyr and M$_{min,Q}{\sim}10^{12}\,{\rm M}_{\odot}$ obtained by \citet{balogh10}. 

The cluster galaxy accretion history models used by all these studies only include model galaxies physically within $r_{200}$ of the halo center.
However, as we estimate that about half of star-forming galaxies with $r_{proj}{<}r_{200}$ spectroscopically identified as cluster members will in fact be physically located outside $r_{200}$ and have never encountered the cluster environment ({\S}~\ref{timescale}), neglecting this contribution could significantly affect their $t_{Q}$ estimates. This is particularly the case for the analysis of the radial population gradients of cluster galaxies by DeL12, who argued that the steep gradients of $f_{pass}$ with cluster-centric radius implied long quenching time-scales (their Fig.~10). As the relative contribution of interloping galaxies physically outside $r_{200}$ increases rapidly with $r_{proj}$ from ${\sim}0$\% at the cluster core to 100\% by 2--3\,$r_{200}$, these steep radial population gradients are in fact best reproduced by rapid or instantaneous quenching ($t_{Q}{=}0$) models (e.g. Fig~7 of Paper I). 

\citet{wetzel} use a similar approach to that presented here to estimate the quenching time-scale as a function of halo mass, by comparing the specific-SFR distributions of satellite and central galaxies. For cluster-mass halos of mass $10^{14}-10^{15}\,{\rm M}_{\odot}$, their data suggested a ``delayed-then-rapid'' form of quenching, in which a satellite can continue to form stars unaffected for ${\sim}3$\,Gyr, before undergoing a rapid quenching with e-folding time-scales of 0.2--0.8\,Gyr. This rather different scenario came from them not finding any significant offset in the peak of the specific-SFR distribution of star-forming galaxies in clusters in comparisons to central (field) galaxies of the same stellar mass \citep{wetzel11}, although their Fig.~2 does hint at a slight offset of order 0.05--0.15\,dex for massive galaxies, similar to that revealed in our Fig.~\ref{ssfr_kmag}. 

It is only by analysing such large samples of cluster galaxies as provided within LoCuSS that we have been able to securely detect the subtle effect of slow quenching on star-forming galaxies within clusters.

\subsection{No triggered star formation in cluster galaxies}

The fraction of lenticulars within groups and clusters has increased rapidly over the last 5\,Gyr, rising from ${\sim}2$0\% in systems at $z{\ga}0.5$ to 50--60\% by the present day \citep{dressler97,treu,wilman}, empirically replacing the star-forming spirals. A key difficulty in explaining this conversion from spirals to S0s since $z{\sim}0.5$ is that S0s differ from spirals by having higher bulge luminosities rather than fainter disks \citep{christlein}. This requires bulge growth during S0 formation, disfavoring a simple fading of the disk component via gas removal mechanisms such as ram-pressure stripping or starvation. 

The simplest method of building up the bulges of cluster lenticulars is to invoke a starburst episode within the central regions of these galaxies, perhaps triggered by galaxy mergers or harassment.
This should manifest itself in the form of significant numbers of starburst galaxies in and around clusters, as nuclear starbursts are triggered in the infall regions of clusters via galaxy interactions or as galaxies encounter the accretion shocks of the cluster's halo. If we identify starbursts as galaxies for which the specific-SFRs are ${\ga}3{\times}$ above the median value of the main sequence of normal star forming field galaxies (${\sim}10^{-10}$\,yr$^{-1}$) at that redshift, then Figure~\ref{ssfr_compare} revealed conclusively that starburst galaxies with specific-SFRs ${\ga}3{\times}10^{-10}$\,yr$^{-1}$ are essentially absent in cluster environments. Indeed, focussing on the high side of the specific-SFR distribution (${\ga}10^{-10}$\,yr$^{-1}$), the specific-SFR distributions of star forming galaxies in our cluster samples appear either entirely consistent with that of field galaxies at the same redshift or even marginally suppressed. 
The specific-SFR distributions of cluster galaxies presented in Fig.~\ref{ssfr_compare} are consistent with \citet{haines11b} who found that the vast majority of star formation in local clusters occurs within quiescent spiral disks, without any significant contribution from nuclear starbursts. 
 
If we require 30--40\% of cluster galaxies to be transformed from spirals to S0s over the last 5\,Gyr via nuclear starbursts, then assuming a 100\,Myr long starburst phase, would imply that ${\sim}0$.6--0.8\% of all cluster galaxies would be observed in this starburst phase at any given time. Assuming initial stellar masses for these galaxies of 0.5--1.$0{\times}10^{11}{\rm M}_{\odot}$ and requiring say 10\% of this stellar mass to be produced via nuclear starburst in order to form the bulge, implies typical SFRs of ${\sim}5$0--100\,M$_{\odot}$\,yr$^{-1}$ over this 100\,Myr starburst period. Over our full sample of 30 systems we have 5\,147 cluster members with $M_{K}{<}{-}23.1$ covered by our 24$\mu$m imaging, and so should expect to see 30--40 such starbursts in our sample, or 1--2 per cluster. We see just four cluster members with SFRs greater than 40\,M$_{\odot}$\,yr$^{-1}$ in our sample (Fig.~\ref{lf}). It seems that if the bulges of the present day cluster S0s were formed in nuclear starbursts this activity must be largely have been confined to galaxy groups, which were later accreted onto the clusters (pre-processing), or have taken place almost entirely in clusters at $z{\ga}0.3$ where at least such starbursts appear frequent \citep[e.g.][]{geach09}. 
This is consistent with \citet{just} who find that the rise in the S0 fraction since $z{\sim}0.5$ is much more rapid in groups and low-$\sigma$ clusters than in massive clusters comparable to those in our sample.

\section{Summary}
\label{sec:summary}

We have presented an analysis of the levels and evolution of star formation activity in a representative sample of 30 massive clusters at $0.15{<}z{<}0.30$ drawn from the Local Cluster Substructure Survey (LoCuSS). We have combined wide-field {\em Spitzer}/MIPS 24$\mu$m data sensitive to obscured SFRs $\sim$3\,M$_{\odot}$\,yr$^{-1}$ in the highest redshift systems with ACReS, our recently completed spectroscopic program, in order to obtain a highly complete census of obscured star formation in the virialized regions of these systems. Our main results are summarized below:

\begin{enumerate}
\item We determined the mid-infrared luminosity function of star-forming cluster galaxies for our full sample of 30 clusters extending down to ${\sim}1{\times}10^{10}L_{\odot}$. We find the infrared LF of cluster galaxies to be indistinguishable to that from a coeval population of field galaxies taken from the same {\em Spitzer} images.
\item We combined our sample of 30 clusters with seven local systems and two clusters at $z{\sim}0.4$ to measure the mid-infrared Butcher--Oemler effect, finding rapid evolution in the fraction of massive ($M_{K}{<}{-}23.1$) cluster galaxies within $1.5\,r_{500}$ with SFRs${>}3$\,M$_{\odot}$\,yr$^{-1}$, of the form \mbox{$f_{SF}{\propto}(1+z)^{7.6{\pm}1.1}$} over the redshift range 0.0--0.4. 

\item We find an even more rapid evolution in the star formation activity of cluster LIRGs, with a ${\sim}15{\times}$ decline in $\Sigma$(SFR)/$M_{200}$ just since $z{\sim}0.3$.  
\item The Butcher--Oemler effect is produced by the combination of a ${\sim}3{\times}$ decline in the mean specific-SFRs of star-forming cluster galaxies since $z{\sim}0.3$ with a ${\sim}1.5{\times}$ decrease in number density. Two-thirds of this evolution in the specific-SFRs of star-forming cluster galaxies is due to the steady cosmic decline in star formation among all galaxies, but one-third reflects an accelerated decline in the star formation activity of cluster galaxies since $z{\sim}0.3$ with respect to that seen in the field. 
\item The specific-SFRs of star-forming cluster galaxies at $z{\sim}0.2$ are found to be systematically ${\sim}28{\pm}3$\% lower than their counterparts in the field at fixed stellar mass and redshift. This is consistent with these galaxies being slowly quenched upon arrival in the cluster, their SFRs declining exponentially on time-scales in the range 0.7--2.0\,Gyr. Such slow quenching is suggestive of ram-pressure stripping or starvation mechanisms. 
\item We find no evidence for a population of starburst galaxies being triggered by accretion into clusters of the kind proposed to transform infalling spirals into present day cluster S0s by building the stellar bulge with a major nuclear starburst episode. 
\end{enumerate}

\section*{Acknowledgements}

CPH, GPS and AJRS acknowledge financial support from STFC.  GPS acknowledges support from the Royal Society. 
We acknowledge NASA funding for this project under the Spitzer program GO:40872.  
This work was supported in part by the National Science Foundation under Grant No. AST-1211349. 
The Millennium Simulation databases used in this paper and the web application providing online access to them were constructed as part of the activities of the German Astrophysical Virtual Observatory.

\appendix

\section{The creation of a fair comparison sample of field galaxies}

As discussed in {\S}\ref{sec:field}, a key requirement in this analysis is the availability of a statistical sample of field galaxies covering the same $0.15{<}z{<}0.30$ redshift range as the 30 clusters in our sample, which are selected in an {\em identical} manner, as well as coming from the same {\em Spitzer} datasets which have uniform coverages from cluster to cluster. As we are attempting to measure rather subtle differences in the star-formation activities of cluster and field galaxies, any systematic mis-match in the stellar-mass or SFR distribution resulting from the selection criteria used to target galaxies for spectroscopy, even at the 10--20\% level, could dramatically impact these comparisons. To minimize any such biases, we have taken great care in selecting field galaxies in narrow redshift slices either side of each cluster, for which the spectroscopic completeness should be indistinguishable from that obtained for the cluster galaxies themselves. There are two primary criteria used to select targets for spectroscopy which limit the redshift range either side of the cluster for which we should remain complete: the $K$-band apparent magnitude limit corresponding to $M_{K}^{*}{+}2$ at the cluster redshift; and the width of the $J{-}K$ color slice about the C-M relation of each cluster used to maximize the observing efficiency of targeting cluster members. We discuss the impact of each one in turn.

\begin{figure*}
\centerline{\includegraphics[width=176mm]{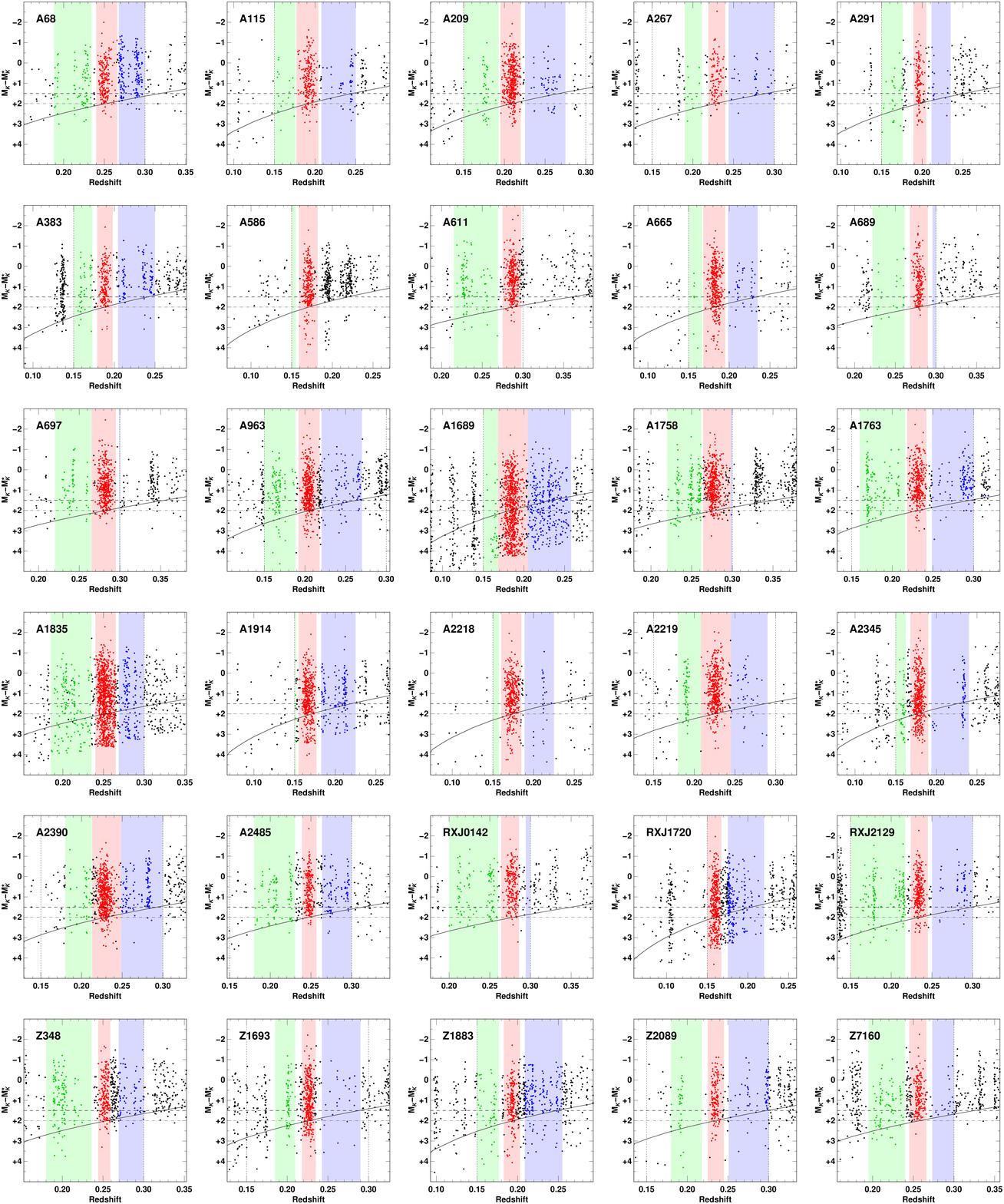}}
\caption{Absolute $K$-band magnitude versus redshift for each galaxy in our spectroscopic sample, with each cluster field shown as a separate panel. The maximal redshift range of the cluster is shown as the red shaded range, while the adjacent redshift ranges used to identify field galaxies in the same dataset are shown by green (foreground field) and blue (background field) shaded regions. The vertical dotted lines indicate the overall $0.15{<}z{<}0.30$ redshift limits used to define our LoCuSS cluster and field galaxy sample. The horizontal dot-dashed lines indicate the overall $M_{K}^{*}{+}2$ limit for selecting cluster members for spectroscopy, with $M_{K}^{*}(z_{cl}){+}2.0$ used to define our faint $K$-band apparent magnitude limit for targeting galaxies for almost all of the clusters. The sloping solid curve shows how this $K$-band apparent magnitude limit produces an effective $M_{K}(z)$ limit which becomes increasingly bright with redshift. The horizontal dashed lines indicate the $M_{K}^{*}{+}1.5$ absolute magnitude limit used throughout this work. }
\label{z_mk}
\end{figure*}

In Figure~\ref{z_mk} we plot $M_{K}$ versus redshift for each galaxy in our spectroscopic sample, with each cluster field shown as a separate panel. The maximal redshift range of the cluster is shown as the red shaded region, while the adjacent redshift ranges used to define our comparison field galaxy sample are indicated by green (foreground field) and blue (background field) shaded regions. Galaxies spectroscopically identifed as belonging to these cluster and field samples are correspondingly shown as red, green or blue symbols, while black points indicate galaxies discarded from this analysis. 
Note the presence of some black points within the cluster redshift range, these are discarded as non-cluster members by virtue of lying outside the cluster caustics. 
There are gaps apparent between the cluster and field redshift ranges for most of the systems, due to our exclusion of galaxies within 4\,000\,km\,s$^{-1}$ of the central cluster redshift from our field sample. The 4\,000\,km\,s$^{-1}$ velocity limit represents the maximal velocity offset for galaxies identified as cluster members, as seen for the most massive systems in our sample (Abell 1689, Abell 1835, Abell 2390), and should ensure that any galaxy in our field sample is at least 50\,Mpc from the cluster. 

Within ACReS we targetted probable cluster galaxies down to a $K$-band apparent magnitude corresponding to $M_{K}^{*}{+}2.0$ ({\em horizontal dot-dashed line}) at the cluster redshift. As we look at field galaxies at redshifts either side of the cluster, this $K$-band apparent magnitude limit produces an effective absolute magnitude limit $M_{K}(z)$ which systematically shifts to brighter values with increasing redshift, as shown by the sloping solid curves. We usually identify our upper redshift limit for selecting field galaxies (right edge of blue shaded region) as the point at which this curve reaches $M_{K}^{*}{+}1.5$ ({\em dashed lines}), or our overall redshift limit of $z{=}0.30$, whichever is lowest. Two systems, Abell 1689 and Abell 1835, have particularly extensive spectroscopy from \citet{czoske} which goes to much fainter magnitude limits than $M_{K}^{*}{+}2.0$, allowing us to extend the redshift range over which we remain complete to $M_{K}^{*}{+}1.5$.

We have explicitly removed redshift ranges from our field sample that contain massive structures, including a $z{=}0.195$ cluster in the same field as Abell 586, and X-ray groups identified from our {\em XMM} imaging, such as the two foreground groups at $z{\sim}0.18$ identified in the fields of Abell 267 and Abell 291. It is notable that even in our ``field'' regions structures and voids in redshift-space are apparent in Fig.~\ref{z_mk}. In the case of Abell 1758 for example, we see clear foreground structures at $z{=}0.23$, $z{=}0.25$ and $z{=}0.26$. We have visually checked the spatial distribution of galaxies from each of these structures on the plane of the sky, and confirm that they are mostly extended filamentary structures, none of which are associated to extended X-ray structures detected in our {\em XMM} images.

In an analogous way, we determined the redshift limits imposed by our $J{-}K$ color cuts by plotting the $J-K$ color offset of each galaxy from the C-M relation versus redshift for each cluster. As the $J{-}K$ color of galaxies increases steadily with redshift (at fixed stellar mass and Hubble type), the vast majority of galaxies are found to lie within a single well-defined diagonal band (${\delta}J{-}K/{\delta}z{\sim}1.0$ at $z{\sim}0.2$) in this plot. The intercept of this diagonal band with our color cuts then defines the redshift range over which our spectroscopic survey should remain complete. The color cuts for each cluster were asymmetically spaced with respect to the C-M relation of that cluster, the upper color cut being located further from the C-M relation than the lower one. The resulting redshift limits are thus also asymmetrically offset from the cluster redshift, with typically the redshift range extending to ${\delta}z{\sim}0.$10--0.15 beyond the cluster, but just ${\sim}0.$05--0.10 to lower redshifts. For each cluster, our final upper redshift limit for our field sample was ultimately defined by the requirement of remaining complete to $M_{K}^{*}{+}1.5$, this being the more stringent requirement than that imposed by the color cuts. The final lower redshift limit for our field sample for each cluster was defined by our lower $J{-}K$ color cut. 

\label{lastpage}
\end{document}